\documentclass[10pt,letter]{article}

\usepackage{fancyhdr}
\usepackage{lastpage}
\usepackage[toc,page]{appendix}
\usepackage[ruled,vlined]{algorithm2e}

\SetKwInput{KwInput}{Input}                
\SetKwInput{KwOutput}{Output}              


\usepackage[parfill]{parskip} 
\usepackage[margin=1in]{geometry}
\usepackage[doublespacing]{setspace}
\usepackage{natbib}
\usepackage{url}
\usepackage{amsfonts,amsmath,bbm,bm,amssymb,mathtools}
\usepackage{xcolor}
\usepackage{ulem}
\usepackage{amsmath}
\usepackage[colorinlistoftodos]{todonotes}
\usepackage{caption} 
\usepackage{multirow}
\usepackage[flushleft]{threeparttable}

\usepackage[pagebackref=true,bookmarks]{hyperref}
\hypersetup{
	unicode=false,
	pdftoolbar=true,
	pdfmenubar=true,
	pdffitwindow=false,     
	pdfstartview={FitH},    
	pdftitle={Penalized GLMM},    
	pdfauthor={JSP},     
	pdfsubject={Subject},   
	pdfcreator={JSP},   
	pdfproducer={JSP}, 
	pdfkeywords={}, 
	pdfnewwindow=true,      
	colorlinks=true,       
	linkcolor=red,          
	citecolor=blue,        
	filecolor=black,      
	urlcolor=blue           
}

\usepackage{xr}
\makeatletter
\newcommand*{\addFileDependency}[1]{
  \typeout{(#1)}
  \@addtofilelist{#1}
  \IfFileExists{#1}{}{\typeout{No file #1.}}
}
\makeatother

\newcommand*{\myexternaldocument}[1]{
    \externaldocument{#1}
    \addFileDependency{#1.tex}
    \addFileDependency{#1.aux}
}

\myexternaldocument{supp}

\def\Var{\text{Var}}

\def\E{\mathbb{E}}

\def\({\left(}
\def\){\right)}
\def\[{\left[}
\def\]{\right]}

\def\glinternet{\texttt{glinternet}}
\def\gesso{\texttt{gesso}}
\def\pglmm{\texttt{pglmm}}

\usepackage{array}
\newcolumntype{H}{>{\setbox0=\hbox\bgroup}c<{\egroup}@{}}

\title{Hierarchical selection of genetic and gene by environment interaction effects in high-dimensional mixed models}

\author{JULIEN ST-PIERRE\thanks{To whom correspondence should be addressed.}\\[4pt]
\textit{Department of Epidemiology, Biostatistics and Occupational Health,} \\
\textit{McGill University, Montreal, Quebec, Canada}
\\[2pt]
{julien.st-pierre@mail.mcgill.ca}\\[4pt]
KARIM OUALKACHA\\[4pt]
\textit{Département de Mathématiques,} \\
\textit{Université du Québec à Montréal, Montreal, Quebec, Canada}\\[4pt]
SAHIR RAI BHATNAGAR\\[4pt]
\textit{Department of Epidemiology, Biostatistics and Occupational Health,} \\
\textit{McGill University, Montreal, Quebec, Canada}}

\begin{document}
\pagestyle{fancy}
\maketitle

\begin{abstract}
Interactions between genes and environmental factors may play a key role in the etiology of many
common disorders that are subject to both genetic and environmental risk factors. Several regularized generalized linear models (GLMs) have been proposed for hierarchical selection of gene by environment interaction (GEI) effects, where a GEI effect is selected only if the corresponding genetic main effect is also selected in the model. However, none of these methods allow to include random effects to account for population structure, subject relatedness and shared environmental exposure. Indeed, the joint estimation of variance components and fixed effects in regularized models is challenging both from a computational and analytical point of view, as the marginal likelihood for a generalized linear mixed model (GLMM) has no closed form except with normal responses. To address these challenges, penalized quasi-likelihood (PQL) estimation is conceptually attractive as under this method, random effects can be treated as fixed effects, which allows to perform regularized estimation of both fixed and random effects in a similar fashion to GLMs. In this paper, we develop a unified approach based on regularized PQL estimation to perform hierarchical selection of GEI effects in sparse regularized mixed models. We propose to use two kinship matrices to account for the fact that individuals can be genetically related while sharing the same environmental exposure. We develop a proximal Newton-type algorithm with block coordinate descent for PQL estimation with mixed lasso and group lasso penalties. We compare the selection and prediction accuracy of our proposed model with existing methods through simulations under the presence of population structure and shared environmental exposure. We show that for all simulation scenarios, compared to other penalized methods, our proposed method enforced sparsity by controlling the number of false positives in the model while having the best predictive performance. Finally, we apply our method to a real data application using the Orofacial Pain: Prospective Evaluation and Risk Assessment (OPPERA) study, and found that our method retrieves previously reported significant loci. Our Julia package \texttt{PenalizedGLMM.jl} is publicly available on github : \url{https://github.com/julstpierre/PenalizedGLMM}.
\end{abstract}

\section{Introduction}
Interactions between genes and environmental factors may play a key role in the multifactorial etiology of many complex diseases that are subject to both genetic and environmental risk factors. For example, in assessing interactions between a polygenic risk score (PRS) and non-genetic risk factors for young-onset breast cancers (YOBC), \citet{Shi2020} showed a decreased association between the PRS and YOBC risk for women who had ever used hormonal birth control, suggesting that environmental exposure might result in risk stratification by interacting with genetic factors.
Thus, there is a raising interest for discovering gene-environment interaction (GEI) effects as they are fundamental to better understand the effect of environmental factors in disease and to increase risk prediction~\citep{Mukherjee2009}.
Several regularized generalized linear models (GLMs) have been proposed for selection of both genetic and GEI effects in genetic association studies~\citep{Fang2023, Zemlianskaia2022,Lim2015}, but currently no such method allows to include any random effect to account for genetic similarity between subjects. Indeed, one can control for population structure and/or closer relatedness by including in the model a polygenic random effect with variance-covariance structure proportional to a kinship or genetic similarity matrix (GSM)~\citep{Yu2005}. However, because kinship is a high-dimensional process, it cannot be fully captured by including only a few Principal Components (PCs) as fixed effects in the model~\citep{Hoffman2013}. Hence, while both Principal Component Analysis (PCA) and mixed models (MMs) share the same underlying model, MMs are more robust in the sense that they do not require distinguishing between the different types of confounders~\citep{Price2010}. Moreover, MMs alleviate the need to evaluate the optimal number of PCs to retain in the model as fixed effects.


Except for normal responses, the joint estimation of variance components and fixed effects in regularized models is challenging both from a computational and analytical point of view, as the marginal likelihood for a generalized linear mixed model (GLMM) has no closed form. To address these challenges, penalized quasi-likelihood (PQL) estimation is conceptually attractive as under this method, random effects can be treated as fixed effects, which allows to perform regularized estimation of both fixed and random effects as in the GLM framework. Several authors have proposed to combine PQL estimation in presence of sparsity by inducing regularization to perform joint selection of fixed and/or random effects in multivariable GLMMs~\citep{StPierre2023,Hu2019,huiJointSelectionMixed2017}. However, these methods were not developed to specifically address selection of GEI effects. Although it is possible to perform naive selection of fixed and GEI effects by simply considering interaction terms as additional predictors, the aforementioned methods are not tailored to perform hierarchical selection, where interaction terms are only allowed to be selected if their corresponding main effects are active in the model~\citep{Bien2013}. Hierarchical variable selection of GEI effects is appealing both for increasing statistical power~\citep{Cox1984} and for enhancing model interpretability because interaction terms that have large main effects are more likely to be retained in the model. 

Population structure and closer relatedness may also cause dependence between gene and environment, leading to selection of spurious GEI effects~\citep{Dudbridge2014}. In the context of genome-wide association studies (GWAS),~\citet{sul2016} showed that under the polygenic model, ignoring this dependence may largely increase the false positive rate of GEI statistics. They proposed introducing an additional random effect that captures the similarity of individuals due to polygenic GEI effects to account for the fact that individuals who are genetically related and who share a common environmental exposure are more closely related. To our knowledge, the spurious selection of GEI effects in regularized models due to the dependence between gene and shared environmental exposure has not been explored yet. Thus, further work is needed to develop sparse regularized GLMMs for hierarchical selection of GEI effects in genetic association studies, while explicitly accounting for the complex correlation structure between individuals that arises from both genetic and environmental factors.

In this paper, we develop a unified approach based on regularized PQL estimation to perform hierarchical selection of GEI effects in sparse regularized logistic mixed models. We rely on the work of~\citet{sul2016} and use two random effects to capture population structure, closer relatedness and shared environmental exposure. We propose to use a composite absolute penalty (CAP) for hierarchical variable selection~\citep{Zhao2009} to seek a sparse subset of genetic and GEI effects that gives an adequate fit to the data. We derive a proximal Newton-type algorithm with block coordinate descent for PQL estimation with mixed lasso and group lasso penalties, relying on our previous work to address computational challenges associated with regularized PQL estimation in high-dimensional data~\citep{StPierre2023}. We compare the prediction and selection accuracy of our proposed model with existing methods through simulations under the presence of population structure and environmental exposure. Finally, we also apply our method to a real data application using the Orofacial Pain: Prospective Evaluation and Risk Assessment (OPPERA) study cohort~\citep{Maixner2011} to study the sex-specific association between temporomandibular disorder (TMD) and genetic predictors.

\section{Methodology}
\subsection{Model}
We have the following generalized linear mixed model (GLMM)
\begin{align}\label{eq:model}
g(\mu_i) &= \eta_i = \bm{Z_i} \bm\theta + D_i\alpha + \bm{G_i} \bm\beta  + (D_i \odot \bm{G}_i) \bm{\gamma}+ b_i
\end{align}
for $i=1,..,n$, where $\mu_i=\E(y_{i}| \bm{Z}_i,\bm{G}_i, D_i, b_i)$, $\bm{Z}_i$ is a $1\times m$ row vector of covariates for subject $i$,  $\bm{G}_i$ is a $1 \times p$ row vector of genotypes for subject $i$ taking values $\{0,1,2\}$ as the number of copies of the minor allele, $(\bm\theta^\intercal,\bm\beta^\intercal)^\intercal$ is a $(m + p) \times 1$ column vector of fixed covariate and additive genotype effects including the intercept, $D_i$ is the exposure of individual $i$ to a binary or continous environmental factor $D$ with fixed effect $\alpha$, $\odot$ is the Hadamard (element-wise) product, and $\bm{\gamma}=[\bm{\gamma}_1,\bm{\gamma}_2,...,\bm{\gamma}_p]^\intercal \in \mathbbm{R}^p$ is the vector of fixed GEI effects. Thus, we have a total of $2p+m+1$ coefficients. We assume that $\bm{b}=(b_1,...,b_n)^\intercal \sim \mathcal{N}(0, \tau_g \bm{K} + \tau_d \bm{K}^D)$ is an $n \times 1$ column vector of random effects, with $\bm{\tau}=(\tau_g,\tau_d)^\intercal$ the variance components that account for the relatedness between individuals. $\bm{K}$ is a known GSM or kinship matrix and $\bm{K}^D$ is an additional kinship matrix that describes how individuals are related both genetically and environmentally, because a pair of individuals who are genetically related and share the same environment exposure have a non-zero kinship coefficient. For a binary exposure, we define $K_{ij}^D=K_{ij}$ if $D_i=D_j$, and $K_{ij}^D=0$ otherwise. For a continuous exposure, one possibility is to set $K_{ij}^D=K_{ij}(1-d(D_i, D_j))$, where $d$ is a metric with range $[0,1]$. The phenotypes $y_i$'s are assumed to be conditionally independent and identically distributed given $(\bm{Z}_i, \bm{G}_i, D_i, \bm{b})$ and follow any exponential family distribution with canonical link function $g(\cdot)$, mean $\E(y_i | \bm{Z}_i, \bm{G}_i, D_i, \bm{b}) =\mu_i$ and variance $\Var(y_i| \bm{Z}_i, \bm{G}_i, D_i, \bm{b}) = \phi a_i^{-1} \nu(\mu_i),$ where $\phi$ is a dispersion parameter, $a_i$ are known weights and $\nu(\cdot)$ is the variance function. 

\subsection{Regularized PQL Estimation}
In order to estimate the model parameters and perform variable selection, we use an approximation method to obtain an analytical closed form for the marginal likelihood of model \eqref{eq:model}. We propose to fit \eqref{eq:model} using a PQL method~\citep{StPierre2023,Chen2016}, from where the log integrated quasi-likelihood function is equal to
\begin{align}\label{eq:A3}
\ell_{PQL}(\bm{\Theta}, \phi, \bm{\tau};\bm{\tilde{b}}) = -\frac{1}{2}\text{log}\left|\left(\tau_g \bm{K} + \tau_d \bm{K}^D\right)\bm{W} + \bm{I}_n\right| + \sum_{i=1}^n ql_i(\bm{\Theta};\bm{\tilde{b}}) - \frac{1}{2}\bm{\tilde{b}}^\intercal \left(\tau_g \bm{K} + \tau_d \bm{K}^D\right)^{-1}\bm{\tilde{b}},
\end{align}
where $\bm{\Theta}=\left(\bm{\theta}^\intercal, \alpha, \bm{\beta}^\intercal,\bm{\gamma}^\intercal\right)^\intercal, \bm{W} = \phi^{-1}\bm{\Delta}^{-1}=\phi^{-1}\textrm{diag}\left\{ \frac{a_i}{\nu(\mu_i)[g'(\mu_i)^2]}\right\}$ is a diagonal matrix containing weights for each observation, $ql_i(\bm{\Theta};\bm{b}) = \int_{y_i}^{\mu_i}\frac{a_i(y_i-\mu)}{\phi\nu(\mu)} d\mu$ is the quasi-likelihood for the $ith$ individual given the random effects $\bm b$, and $\tilde{\bm{b}}$ is the solution which maximizes $\sum_{i=1}^n ql_i(\bm{\Theta};\bm{{b}}) - \frac{1}{2}\bm{{b}}^\intercal \left(\tau_g \bm{K} + \tau_d \bm{K}^D\right)^{-1}\bm{{b}}$.

In typical genome-wide studies, the number of genetic predictors is much greater than the number of observations ($p > n$), and the fixed effects parameter vector $\bm{\Theta}$ becomes underdetermined when modelling $p$ SNPs jointly. Moreover, we would like to induce a hierarchical structure, that is, a GEI effect can be present only if both exposure and genetic main effects are also included in the model. Thus, we propose to add a composite absolute penalty (CAP) for hierarchical variable selection~\citep{Zhao2009} to the negative quasi-likelihood function in \eqref{eq:A3} to seek a sparse subset of genetic and GEI effects that gives an adequate fit to the data. We define the following objective function $Q_{\lambda}$ which we seek to minimize with respect to $(\bm{\Theta}, \phi, \bm{\tau})$:
\begin{align}\label{eq:objfunc}
Q_{\lambda}(\bm{\Theta}, \phi, \bm{\tau};\bm{\tilde{b}}) := -\ell_{PQL}(\bm{\Theta}, \phi, \bm{\tau};\bm{\tilde{b}}) +(1-\rho)\lambda\sum_j \|\beta_j, \gamma_j\|_2 + \rho\lambda\sum_j|\gamma_j|,
\end{align}
where $\lambda>0$ controls the strength of the overall regularization and $\rho \in [0,1)$ controls the relative sparsity of the GEI effects for each SNP. In our modelling approach, we do not penalize the environmental exposure fixed effect $\alpha$. Thus, a value of $\rho=0$ is equivalent to a group lasso penalty where we only include a predictor in the model if both its main effect $\beta_j$ and GEI effect $\gamma_j$ are non-zero. A value of $0<\rho<1$ is equivalent to a sparse group lasso penalty where main effects can be selected without their corresponding GEI effects due to the different strengths of penalization, but a GEI effect is still only included in the model if the corresponding main effect is non-zero. 

\subsection{Estimation of variance components}
Jointly estimating the variance components $\tau_g, \tau_d$ and scale parameter $\phi$ with the regression effects vector $\bm{\Theta}$ and random effects vector $\bm{b}$ is a computationally challenging non-convex optimization problem. Updates for $\tau_g, \tau_d$ and $\phi$ based on a majorization-minimization algorithm~\citep{Zhou2019} would require inverting three different $n\times n$ matrices, with complexity $O(n^3)$, at each iteration. Thus, even for moderately small sample sizes, this is not practicable for genome-wide studies. Instead, as detailed in~\citet{StPierre2023}, we propose a two-step method where variance components and scale parameter are estimated only once under the null association of no genetic effect, that is assuming $\bm{\beta}=\bm{\gamma}=0$, using the AI-REML algorithm~\citep{Gilmour1995}. 

\subsection{Spectral decomposition of the random effects covariance matrix}
Given $\hat\tau_g, \hat\tau_d$ and  $\hat\phi$ estimated under the null, spectral decomposition of the random effects covariance matrix yields
 \begin{align}\label{eq:eigen}
 \left(\hat\tau_g \bm{K} + \hat\tau_d \bm{K}^D\right)^{-1} & = \left(\bm{U\Lambda U}^\intercal\right)^{-1} \nonumber \\
 &= \bm{U}\bm{\Lambda}^{-1}\bm{U}^\intercal,
 \end{align}
 where $\bm{U}$ is an orthonormal matrix of eigenvectors and $\bm \Lambda$ is a diagonal matrix of eigenvalues $\Lambda_1\ge \Lambda_2 \ge...\ge \Lambda_n > 0$ because both $\bm{K}$ and $\bm{K}^D$ are positive definite. 
 
 Using \eqref{eq:eigen} and assuming that the weights in $\bm{W}$ vary slowly with the conditional mean~\citep{breslowApproximateInferenceGeneralized1993}, minimizing \eqref{eq:objfunc} is now equivalent to
\begin{align}\label{eq:objfunc2}
\hat{\bm{\Theta}} &=\underset{\bm{\Theta}}{\text{argmin }} -\sum_{i=1}^n ql_i(\bm{\Theta};\bm{\tilde{\delta}}) + \frac{1}{2}\bm{\tilde{\delta}}^\intercal\bm{\Lambda}^{-1}\bm{\tilde{\delta}} + (1-\rho)\lambda\sum_j \|\beta_j, \gamma_j\|_2 + \rho\lambda\sum_j|\gamma_j| \nonumber \\
&= \underset{\bm{\Theta}}{\text{argmin }} f(\bm{\Theta};\bm{\tilde{\delta}}) + g(\bm{\Theta}),
\end{align}
where $\bm{\tilde{\delta}} = \bm{U}^\intercal\bm{\tilde{b}}$ is the minimizer of $f(\bm{\Theta};\bm{{\delta}}) := -\sum_{i=1}^n ql_i(\bm{\Theta};\bm{{\delta}}) + \frac{1}{2}\bm{{\delta}}^\intercal\bm{\Lambda}^{-1}\bm{{\delta}}$. Thus, iteratively solving \eqref{eq:objfunc2} also requires updating the solution $\bm{\tilde{\delta}}$ at each step until convergence. Conditioning on the previous solution for $\bm{\Theta}$, $\bm{\tilde{\delta}}$ is obtained by minimizing a generalized ridge weighted least-squares (WLS) problem with $\bm{\Lambda}^{-1}$ as the regularization matrix. Then, conditioning on $\bm{\tilde{\delta}}$, $\hat{\bm{\Theta}}$ is found by minimizing a WLS problem with a sparse group lasso penalty. We present in Appendix \ref{appendix:1} our proposed proximal Newton-type algorithm that cycles through updates of $\tilde{\bm{\delta}}$ and $\bm{\Theta}$.

\section{Simulation study}
We first evaluated the performance of our proposed method, called \texttt{pglmm}, against that of a standard logistic lasso, using the \texttt{Julia} package \texttt{GLMNet} which wraps the \texttt{Fortran} code from the original \texttt{R} package \texttt{glmnet}~\citep{glmnet}. Then, among logistic models that impose hierarchical interactions, we compared our method with the \texttt{glinternet}~\citep{Lim2015} and \texttt{gesso}~\citep{Zemlianskaia2022} models which are both implemented in \texttt{R} packages. The \texttt{glinternet} method relies on overlapping group lasso, and even though it is optimized for selection of gene by gene interactions in high-dimensional data, it is applicable for selection of GEI effects. An advantage of the method is that it only requires tuning a single parameter value. On the other hand, \texttt{gesso} uses a sparse group lasso formulation to induce a hierarchical structure with a group $L_\infty$ penalty, and the default implementation fits solutions paths across a two-dimensional grid of tuning parameter values. For all methods, selection of the tuning parameters is performed by cross-validation. Finally, for \texttt{glmnet}, \texttt{gesso} and \texttt{glinternet}, population structure and environmental exposure is accounted for by adding the top 10 PCs of the kinship matrix as additional covariates.

\begin{table}[h]
\caption{Number of samples by population for the high quality harmonized set of 4,097 whole genomes from the Human Genome Diversity Project (HGDP) and the 1000 Genomes Project (1000G).}\label{tab:pop}
\centering
\begin{tabular}{lccl}
  \hline
  Population & 1000 Genomes & HGDP & Total \\ 
  \hline
  African & 879 (28\%) & 110 (12\%) & 989 (24\%) \\ 
  Admixed American & 487 (15\%) &  62 (7\%) & 549 (13\%) \\ 
  Central/South Asian & 599 (19\%) & 184 (20\%) & 783 (19\%) \\ 
  East Asian & 583 (18\%) & 234 (25\%) & 817 (20\%) \\ 
  European & 618 (20\%) & 153 (16\%) & 771 (19\%) \\ 
  Middle Eastern &   0 & 158 (17\%) & 158 (4\%) \\ 
  Oceanian &   0 &  30 (3\%) &  30 (1\%) \\ \hline 
  Total & 3166 & 931 & 4097 \\ 
   \hline
\end{tabular}
\end{table}

\subsection{Simulation model}

We performed a total of 100 replications for each of our simulation scenarios, drawing anew genotypes and simulated traits, using real genotype data from a high quality harmonized set of 4,097 whole genomes from the Human Genome Diversity Project (HGDP) and the 1000 Genomes Project (1000G)~\citep{Koenig2023}. At each replication, we sampled {$10\ 000$} candidate SNPs from the chromosome 21 and randomly selected $1\%$ to be causal. Let $S$ be the set of candidate causal SNPs, with $|S|=100$, then the causal SNPs fixed effects $\beta_j$ were generated from a Gaussian distribution $\mathcal{N}(0,h^2_S\sigma^2/|S|)$, where $h^2_S$ is the fraction of variance on the logit scale that is due to total additive genetic fixed effects. Let $S'$ be the set of candidate causal SNPs, not necessarily overlapping with $S$, that have a non-zero GEI effect, with $|S'|=50$, then the GEI effects $\gamma_j$ were generated from a Gaussian distribution $\mathcal{N}(0,h^2_{S'}\sigma^2/|S'|)$, where $h^2_{S'}$ is the fraction of variance on the logit scale that is due to total additive GEI fixed effects. 
Further, we simulated a random effect from a Gaussian distribution $\epsilon\sim\mathcal{N}(0, h^2_{g}\sigma^2\bm{K} + h^2_{d}\sigma^2\bm{K^D})$, where $h^2_{g}$ and $h^2_{d}$ are the fractions of variance explained by the polygenic and polygenic by environment effects respectively. The kinship matrices $\bm{K}$ and $\bm{K^D}$ were calculated using a set of $50,000$ randomly sampled SNPs excluding the set of candidate SNPs, and PCs were obtained from the singular value decomposition of $\bm{K}$. We simulated a covariate for age using a Normal distribution and used the sex covariate provided with the data as a proxy for environmental exposure. Then, binary phenotypes were generated using the following model
\begin{align}\label{eq:simlogit1}
\text{logit}(\pi) = \text{logit}(\pi_{0k}) -\text{log}(1.3)\times Sex+\text{log}(1.05)Age/10 + \sum_{j\in S}\beta_j\cdot \widetilde{G}_{j} + \sum_{j\in S'}\gamma_j\cdot (Sex \odot \widetilde{G}_{j}) + \epsilon,
\end{align}
where $\pi_{0k}$, for $k=1,...,7$, was simulated using a $U(0.1, 0.9)$ distribution to specify a different prevalence for each population in Table \ref{tab:pop} under the null, and $\widetilde{G}_{j}$ is the $j^{th}$ column of the standardized genotype matrix $\tilde{g}_{ij}=(g_{ij}-2p_i)/\sqrt{2p_i(1-p_i)}$ and $p_j$ is the minor allele frequency (MAF) for the $j^{th}$ predictor. 

In all simulation scenarios, we evaluated the methods when $h^2_{g}=0.2$ and $h^2_d=0$ (i.e., low polygenic effects), and when $h^2_{g}=0.4$ and $h^2_d=0.2$ (i.e., high polygenic effects) respectively. In the first simulation scenario, we set $h^2_S=0.2$ and $h^2_{S'}=0.1$ such that each main effect or GEI effect explains 0.2\% of the total variability on the logit scale, and we induced a hierarchical structure for the simulated data by imposing $\gamma_j \ne 0 \rightarrow \beta_j \ne 0$ for $j=1,...,p$. In the second simulation scenario, we repeated the simulations from the first scenario, but without enforcing any hierarchical structure. 

\subsection{Results}

\begin{figure}[t]
\centering
\caption{Precision of compared methods averaged over 100 replications as a function of the number of active GEI effects.}
\includegraphics[scale=0.85]{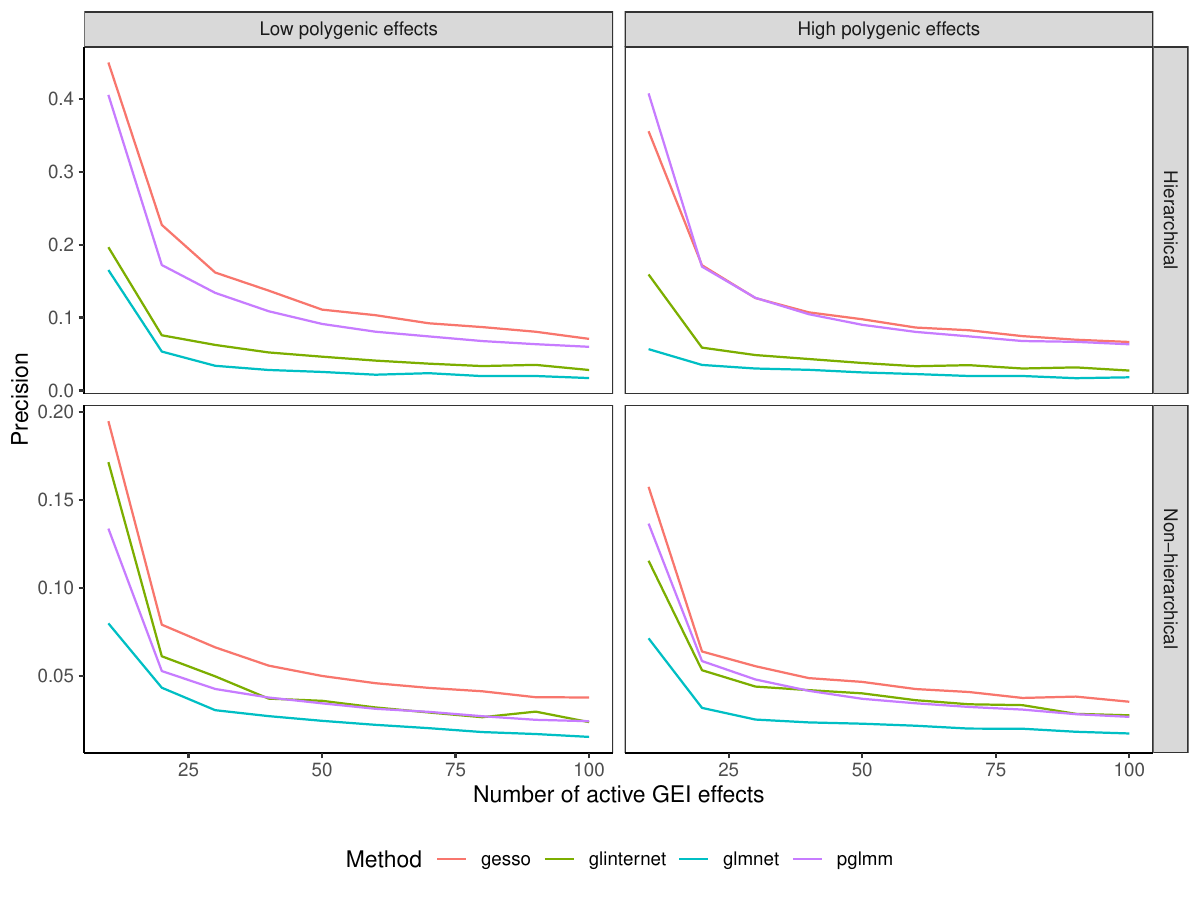}
\label{fig:1}
\end{figure}

\begin{figure}[t]
\centering
\caption{Precision of compared methods averaged over 100 replications as a function of the number of active main effects in the model.}
\includegraphics[scale=0.85]{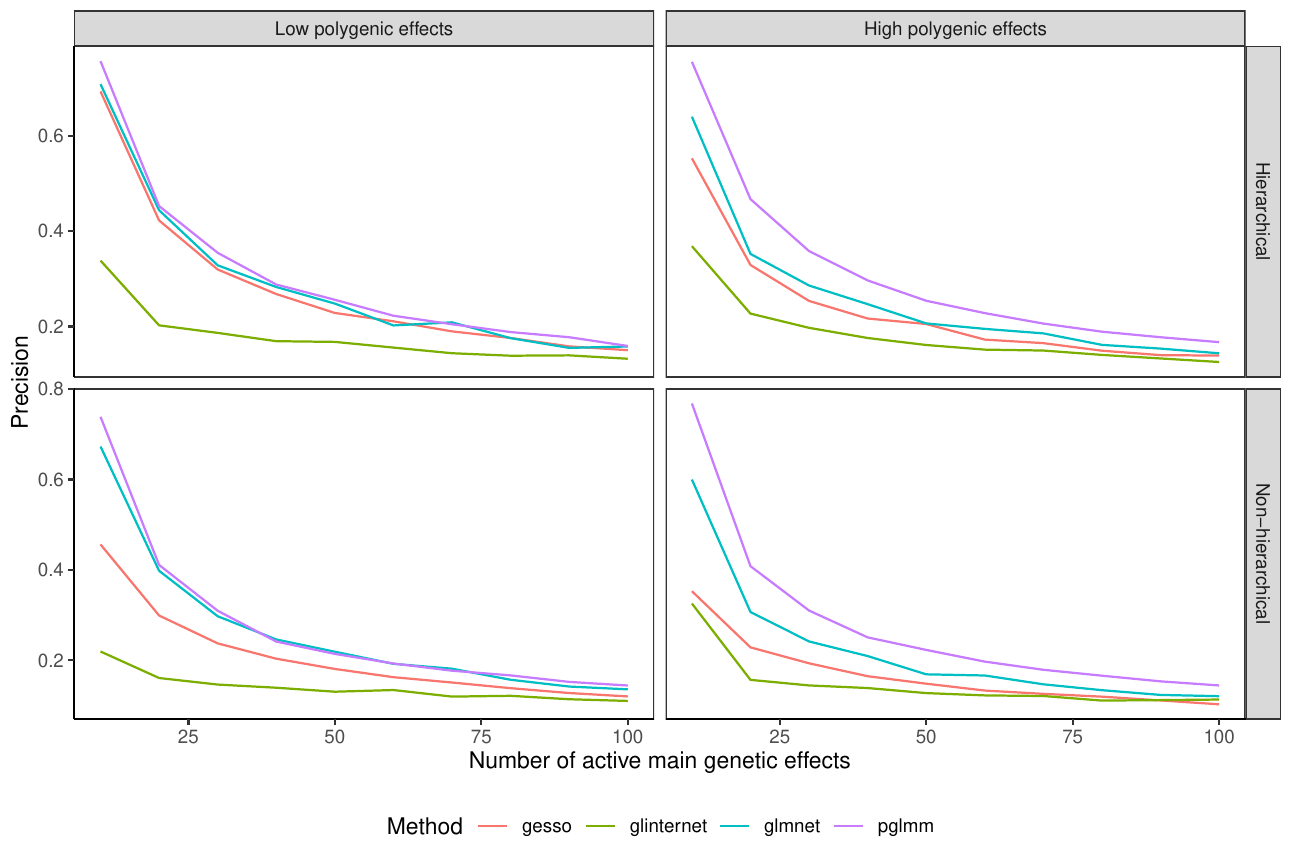}
\label{fig:2}
\end{figure}

We obtained solutions paths across a one dimensional (\texttt{glmnet, glinternet}) or two-dimensional grid of tuning parameter values (\texttt{gesso}, \texttt{pglmm}) for the hierarchical and non-hierarchical simulation scenarios and reported the mean precision, i.e. the proportion of selected predictors that are causal, over 100 replications for the selection of GEI effects (Figure \ref{fig:1}) and main genetic effects (Figure \ref{fig:2}) respectively. We see from Figure \ref{fig:1} that in the hierarchical simulation scenario, \texttt{gesso} and \texttt{pglmm} retrieve important GEI effects with better precision than \texttt{glmnet} and \texttt{glinternet}. When we simulate no random polygenic GEI effect, \texttt{gesso} slightly outperforms \texttt{pglmm}, but when we increase the heritability of the two random effects, both methods perform similarly. When we simulate data under no hierarchical assumption, precision for all hierarchical models fall drastically, although they still perform better than the standard lasso model. We note that \texttt{gesso} retrieves important GEI effects with equal or better precision than other methods in all simulation settings. On the other hand, we see from Figure \ref{fig:2} that \texttt{pglmm} outperforms all methods for retrieving important main effects for both hierarchical and non-hierarchical simulation scenarios. When we simulate low polygenic effects, \texttt{pglmm} and \texttt{glmnet} perform comparably. We also note that \texttt{gesso} retrieves main effects with less precision than \texttt{glmnet} and \texttt{pglmm} in all scenarios. At last, the precision of \texttt{glinternet} is considerably lower than all other methods until the number of active main genetic effects in the model is large.

In practice, we often do not have any a priori knowledge for the number of main effects and/or GEI effects that we want to include in the final model. Thus, instead of comparing methods at a fixed number of active predictors, we can use cross-validation as a model selection criteria. For the two simulation scenarios that we previously described, we randomly split the data into training and test subjects, using a 80/20 ratio, and fitted the full lasso solution path on the training set for 100 replications. We report the model size, false positive rate (FPR), true positive rate (TPR), false discovery rate (FDR), and $F_1$ score on the training sets, and the area under the ROC curve (AUC) when making predictions on the independent test subjects. The $F_1$ score, which is defined as the harmonic mean of the precision ($1-FDR$) and TPR, can be used to take into account that methods with a large number of selected predictors will likely have a higher TPR, and inversely that methods with a lower number of selected predictors will likely have a higher precision. Results for the selection of main genetic effects, GEI effects and combined main and GEI effects are included in Table \ref{tab:GEI}, Table \ref{tab:main} and Table \ref{tab:all} respectively. 

With respect to selection of the GEI effects, the comparative performance of each method varies depending on the simulation scenario. In the hierarchical scenario, \texttt{pglmm} has the highest $F_1$ score for both low and high heritability of the random effects, while in the anti-hierarchical scenario, \gesso~has the highest TPR and $F_1$ score. When heritability of the polygenic random effects is high, \pglmm~has the lowest FDR. With respect to the genetic main effects, \texttt{pglmm} selects the lowest number of predictors in the model, and thus has the lowest FPR and FDR in all simulation scenarios. On the other hand, \texttt{glinternet} always selects the largest number of predictors in all scenarios, and hence has the highest TPR. Using the $F_1$ score to balance FDR and TPR, we see that \texttt{pglmm} performs the best for retrieving the important main genetic effects in all simulation scenarios. Also, we see that \texttt{gesso} and \texttt{pglmm} perform similarly when the heritability of the polygenic random effect is low, but when we increase the heritability and add a second random effect for the GEI, the FDR for \gesso~increases drastically, and the number of selected main effects becomes on average more than two times higher than for \texttt{pglmm}.  Finally, when we compare the selection performance of all methods for the combined genetic main effects and GEI effects, we have that for all simulation scenarios, \pglmm~always selects the lowest number of predictors in the model, has the lowest FPR and FDR, and has the highest prediction accuracy in independent subjects as measured by the AUC. Thus, compared to other methods, \pglmm~enforces sparsity by controlling the number of false positives in the model while having the best predictive performance, at the cost of reducing the number of true positives. By using the $F_1$ score to account for this trade-off between FDR and TPR, giving equal importance to both, we have that \pglmm~performs the best in term of selection of important predictors in all simulation scenarios.

\begin{table}[ht]
\centering
\begin{threeparttable}

\caption{Results for the GEI effects $\gamma$. For each simulation scenario, we report the mean value over 100 replications when we simulate only one random effect with low heritability (Low $\epsilon$) and we simulate two random effects with high heritability (High $\epsilon$).}\label{tab:GEI}
\begin{tabular}{llHHcccc}
    \hline
   & & & & \multicolumn{2}{c}{Non-hierarchical model} & \multicolumn{2}{c}{Hierarchical model} \\
  \hline Metric & Method & Low $\epsilon$ & High $\epsilon$ & Low $\epsilon$ & High $\epsilon$ & Low $\epsilon$ & High $\epsilon$ \\
  \hline
Model size & \texttt{glmnet} & 1.7 & 19.2 & 13.5 & 39.7 & 13.7 & 38.9 \\ 
   & \glinternet & 3.78 & 17.5 & 10.7 & 39.9 & 10.3 & 39.8 \\ 
   & \gesso & 9.15 & 63.3 & 45 & 105 & 60 & 117 \\ 
   & \pglmm & 12.4 & 10 & 49.1 & 43.1 & 54 & 48.9 \\ 
   \hline
FPR & \texttt{glmnet} & \bf{1.70}$\times\bf{10^{-4}}$ & 1.92$\times10^{-3}$ & 1.29$\times10^{-3}$ & 3.90$\times10^{-3}$ & 1.31$\times10^{-3}$ & \bf{3.82}$\times\bf{10^{-3}}$ \\ 
   & \glinternet & 3.78$\times10^{-4}$ & 1.75$\times10^{-3}$ & \bf{9.97}$\times\bf{10^{-4}}$ & \bf{3.85}$\times\bf{10^{-3}}$ & \bf{9.55}$\times\bf{10^{-4}}$ & 3.85$\times10^{-3}$ \\ 
   & \gesso & 9.15$\times10^{-4}$ & 6.33$\times10^{-3}$ & 4.31$\times10^{-3}$ & 1.02$\times10^{-2}$ & 5.60$\times10^{-3}$ & 1.12$\times10^{-2}$ \\ 
   & \pglmm & 1.24$\times10^{-3}$ & \bf{1.00}$\times\bf{10^{-3}}$ & 4.78$\times10^{-3}$ & 4.19$\times10^{-3}$ & 4.99$\times10^{-3}$ & 4.51$\times10^{-3}$ \\ 
   \hline
TPR & \texttt{glmnet} & 0 & 0 & 0.012 & 0.018 & 0.013 & 0.019 \\ 
   & \glinternet & 0 & 0 & 0.015 & 0.033 & 0.017 & 0.030 \\ 
   & \gesso & 0 & 0 & \bf{0.043} & \bf{0.075} & 0.085 & \bf{0.108} \\ 
   & \pglmm & 0 & 0 & 0.031 & 0.029 & \bf{0.087} & 0.081 \\ 
   \hline
FDR & \texttt{glmnet} & 1 & 1 & 0.957 & 0.976 & 0.949 & 0.975 \\ 
   & \glinternet & 1 & 1 & \bf{0.910} & 0.959 & \bf{0.888} & 0.961 \\ 
   & \gesso & 1 & 1 & 0.913 & 0.960 & \bf{0.888} & 0.945 \\ 
   & \pglmm & 1 & 1 & 0.943 & \bf{0.920} & 0.900 & \bf{0.869} \\ 
   \hline
$F_1$ & \texttt{glmnet} & 0 & 0 & 0.043 & 0.032 & 0.046 & 0.035 \\ 
   & \glinternet & 0 & 0 & 0.047 & 0.045 & 0.047 & 0.043 \\ 
   & \gesso & 0 & 0 & \bf{0.056} & \bf{0.052} & 0.084 & 0.067 \\ 
   & \pglmm & 0 & 0 & 0.041 & 0.042 & \bf{0.095} & \bf{0.096} \\ 
   \hline
  \end{tabular}
\begin{tablenotes}
      \item Model size is defined as $\sum_{j=1}^p\mathbbm{I}(\hat{\gamma}_j \ne 0)$.
      \item FPR is defined as $\sum_{j=1}^p\mathbbm{I}(\hat{\gamma}_j \ne 0 \ \cap \gamma_j = 0) / \sum_{j=1}^p\mathbbm{I}(\gamma_j = 0)$.
      \item TPR is defined as $\sum_{j=1}^p\mathbbm{I}(\hat{\gamma}_j \ne 0 \ \cap \gamma_j \ne 0) / \sum_{j=1}^p\mathbbm{I}(\gamma_j \ne 0)$.
      \item FDR is defined as $\sum_{j=1}^p\mathbbm{I}(\hat{\gamma}_j \ne 0 \ \cap \gamma_j = 0) / \sum_{j=1}^p\mathbbm{I}(\hat{\gamma}_j \ne 0)$.
      \item $F_1$ is defined as $2\times\left(\frac{1}{1-FDR} + \frac{1}{TPR}\right)^{-1}$.
    \end{tablenotes}
\end{threeparttable}
\end{table}

\begin{table}[ht]
\centering
\begin{threeparttable}
\caption{Results for the genetic predictors main effects $\beta$. For each simulation scenario, we report the mean value over 100 replications when we simulate only one random effect with low heritability (Low $\epsilon$) and when we simulate two random effects with high heritability (High $\epsilon$).}\label{tab:main}
\begin{tabular}{llHHcccc}
\hline
   & & & & \multicolumn{2}{c}{Non-hierarchical model} & \multicolumn{2}{c}{Hierarchical model} \\
  \hline Metric & Method & Low $\epsilon$ & High $\epsilon$ & Low $\epsilon$ & High $\epsilon$ & Low $\epsilon$ & High $\epsilon$ \\
  \hline
Model size & \texttt{glmnet} & 44.6 & 221 & 231 & 437 & 231 & 440 \\ 
   & \glinternet & 126 & 262 & 269 & 487 & 266 & 488 \\ 
   & \gesso & 28.5 & 156 & 166 & 357 & 171 & 355 \\ 
   & \pglmm & 25.4 & 24.7 & 146 & 142 & 156 & 132 \\ 
   \hline
FPR & \texttt{glmnet} & 4.46$\times10^{-3}$ & 2.21$\times10^{-2}$ & 2.15$\times10^{-2}$ & 4.19$\times10^{-2}$ & 2.13$\times10^{-2}$ & 4.19$\times10^{-2}$ \\ 
   & \glinternet & 1.26$\times10^{-2}$ & 2.62$\times10^{-2}$ & 2.53$\times10^{-2}$ & 4.68$\times10^{-2}$ & 2.47$\times10^{-2}$ & 4.66$\times10^{-2}$ \\ 
   & \gesso & 2.85$\times10^{-3}$ & 1.56$\times10^{-2}$ & 1.52$\times10^{-2}$ & 3.40$\times10^{-2}$ & 1.54$\times10^{-2}$ & 3.35$\times10^{-2}$ \\ 
   & \pglmm & \bf{2.54}$\times\bf{10^{-3}}$ & \bf{2.47}$\times\bf{10^{-3}}$ & \bf{1.32}$\times\bf{10^{-2}}$ & \bf{1.28}$\times\bf{10^{-2}}$ & \bf{1.39}$\times\bf{10^{-2}}$ & \bf{1.16}$\times\bf{10^{-2}}$ \\ 
   \hline
TPR & \texttt{glmnet} & 0 & 0 & 0.186 & 0.228 & 0.198 & 0.255 \\ 
   & \glinternet & 0 & 0 & \bf{0.189} & \bf{0.236} & \bf{0.208} & \bf{0.269} \\ 
   & \gesso & 0 & 0 & 0.157 & 0.206 & 0.181 & 0.239 \\ 
   & \pglmm & 0 & 0 & 0.153 & 0.152 & 0.178 & 0.168 \\ 
   \hline
FDR & \texttt{glmnet} & 1 & 1 & 0.913 & 0.947 & 0.907 & 0.941 \\ 
   & \glinternet & 1 & 1 & 0.928 & 0.950 & 0.918 & 0.944 \\ 
   & \gesso & 1 & 1 & 0.893 & 0.940 & 0.881 & 0.930 \\ 
   & \pglmm & 1 & 1 & \bf{0.877} & \bf{0.875} & \bf{0.867} & \bf{0.837} \\ 
   \hline
$F_1$ & \texttt{glmnet} & 0 & 0 & 0.114 & 0.086 & 0.122 & 0.095 \\ 
   & \glinternet & 0 & 0 & 0.102 & 0.082 & 0.116 & 0.092 \\ 
   & \gesso & 0 & 0 & 0.118 & 0.092 & 0.136 & 0.106 \\ 
   & \pglmm & 0 & 0 & \bf{0.127} & \bf{0.127} & \bf{0.140} & \bf{0.147} \\ 
   \hline
  \end{tabular}
\begin{tablenotes}
      \item Model size is defined as $\sum_{j=1}^p\mathbbm{I}(\hat{\beta}_j \ne 0)$.
      \item FPR is defined as $\sum_{j=1}^p\mathbbm{I}(\hat{\beta}_j \ne 0 \ \cap \beta_j = 0) / \sum_{j=1}^p\mathbbm{I}(\beta_j = 0)$.
      \item TPR is defined as $\sum_{j=1}^p\mathbbm{I}(\hat{\beta}_j \ne 0 \ \cap \beta_j \ne 0) / \sum_{j=1}^p\mathbbm{I}(\beta_j \ne 0)$.
      \item FDR is defined as $\sum_{j=1}^p\mathbbm{I}(\hat{\beta}_j \ne 0 \ \cap \beta_j = 0) / \sum_{j=1}^p\mathbbm{I}(\hat{\beta}_j \ne 0)$.
      \item $F_1$ is defined as $2\times\left(\frac{1}{1-FDR} + \frac{1}{TPR}\right)^{-1}$.
    \end{tablenotes}
\end{threeparttable}
\end{table}

\begin{table}[ht]
\centering
\begin{threeparttable}
\caption{Results for the combined genetic predictors main effects $\beta$ and GEI effects $\gamma$. For each simulation scenario, we report the mean value over 100 replications when we simulate only one random effect with low heritability (Low $\epsilon$) and when we simulate two random effects with high heritability (High $\epsilon$).}\label{tab:all}
\begin{tabular}{llHHcccc}
\hline
   & & & & \multicolumn{2}{c}{Non-hierarchical model} & \multicolumn{2}{c}{Hierarchical model} \\
  \hline Metric & Method & Low $\epsilon$ & High $\epsilon$ & Low $\epsilon$ & High $\epsilon$ & Low $\epsilon$ & High $\epsilon$ \\ \hline
Model size & \texttt{glmnet} & 46.3 & 240 & 245 & 477 & 244 & 478 \\ 
   & \glinternet & 130 & 279 & 280 & 527 & 276 & 529 \\ 
   & \gesso & 37.6 & 219 & 211 & 462 & 231 & 484 \\ 
   & \pglmm & 37.8 & 34.8 & 195 & 185 & 209 & 186 \\ 
   \hline
FPR & \texttt{glmnet} & 2.31$\times10^{-3}$ & 1.20$\times10^{-2}$ & 1.14$\times10^{-2}$ & 2.28$\times10^{-2}$ & 1.13$\times10^{-2}$ & 2.28$\times10^{-2}$ \\ 
   & \glinternet & 6.51$\times10^{-3}$ & 1.40$\times10^{-2}$ & 1.31$\times10^{-2}$ & 2.53$\times10^{-2}$ & 1.28$\times10^{-2}$ & 2.52$\times10^{-2}$ \\ 
   & \gesso & \bf{1.88}$\times\bf{10^{-3}}$ & 1.10$\times10^{-2}$ & 9.72$\times10^{-3}$ & 2.21$\times10^{-2}$ & 1.05$\times10^{-2}$ & 2.29$\times10^{-2}$ \\ 
   & \pglmm & 1.89$\times10^{-3}$ & \bf{1.74}$\times\bf{10^{-3}}$ & \bf{8.98}$\times\bf{10^{-3}}$ & \bf{8.50}$\times\bf{10^{-3}}$ & \bf{9.44}$\times\bf{10^{-3}}$ & \bf{8.30}$\times\bf{10^{-3}}$ \\ 
   \hline
TPR & \texttt{glmnet} & 0 & 0 & 0.128 & 0.158 & 0.137 & 0.177 \\ 
   & \glinternet & 0 & 0 & \bf{0.131} & \bf{0.168} & 0.144 & 0.189 \\ 
   & \gesso & 0 & 0 & 0.119 & 0.162 & \bf{0.149} & \bf{0.198} \\ 
   & \pglmm & 0 & 0 & 0.112 & 0.111 & 0.147 & 0.140 \\ 
   \hline
FDR & \texttt{glmnet} & 1 & 1 & 0.915 & 0.949 & 0.909 & 0.943 \\ 
   & \glinternet & 1 & 1 & 0.928 & 0.951 & 0.918 & 0.945 \\ 
   & \gesso & 1 & 1 & 0.903 & 0.945 & 0.887 & 0.936 \\ 
   & \pglmm & 1 & 1 & \bf{0.899} & \bf{0.892} & \bf{0.877} & \bf{0.859} \\ 
   \hline
$F_1$ & \texttt{glmnet} & 0 & 0 & 0.098 & 0.076 & 0.105 & 0.085 \\ 
   & \glinternet & 0 & 0 & 0.091 & 0.075 & 0.103 & 0.084 \\ 
   & \gesso & 0 & 0 & \bf{0.099} & 0.081 & 0.120 & 0.095 \\ 
   & \pglmm & 0 & 0 & \bf{0.099} & \bf{0.100} & \bf{0.123} & \bf{0.125} \\ 
   \hline
AUC & \texttt{glmnet} & 0.696 & 0.693 & 0.709 & 0.758 & 0.712 & 0.751 \\ 
   & \glinternet & 0.693 & 0.694 & 0.726 & 0.763 & 0.734 & 0.762 \\ 
   & \gesso & 0.702 & 0.691 & 0.725 & 0.755 & 0.731 & 0.756 \\ 
   & \pglmm & \bf{0.719} & \bf{0.746} & \bf{0.745} & \bf{0.801} & \bf{0.752} & \bf{0.799} \\ 
   \hline
  \end{tabular}
\begin{tablenotes}
      \item $\bm{\theta}=\{\beta_1,...,\beta_p,\gamma_1,...,\gamma_p\}$.
      \item Model size is defined as $\sum_{j=1}^{2p}\mathbbm{I}(\hat{\theta}_j \ne 0)$.
      \item FPR is defined as $\sum_{j=1}^{2p}\mathbbm{I}(\hat{\theta}_j \ne 0 \ \cap \theta_j = 0) / \sum_{j=1}^{2p}\mathbbm{I}(\theta_j = 0)$.
      \item TPR is defined as $\sum_{j=1}^{2p}\mathbbm{I}(\hat{\theta}_j \ne 0 \ \cap \theta_j \ne 0) / \sum_{j=1}^{2p}\mathbbm{I}(\theta_j \ne 0)$.
      \item FDR is defined as $\sum_{j=1}^{2p}\mathbbm{I}(\hat{\theta}_j \ne 0 \ \cap \theta_j = 0) / \sum_{j=1}^{2p}\mathbbm{I}(\hat{\theta}_j \ne 0)$.
      \item $F_1$ is defined as $2\times\left(\frac{1}{1-FDR} + \frac{1}{TPR}\right)^{-1}$.
    \end{tablenotes}
\end{threeparttable}
\end{table}


\clearpage
\section{Discovering sex-specific genetic predictors of painful temporomandibular disorder}

Significant associations between temporomandibular disorder (TMD), which is a painful disease of the jaw, and four distinct loci have been previously reported in combined or sex-segregated analyses on the Orofacial Pain: Prospective Evaluation and Risk Assessment (OPPERA) study cohort \citep{Smith2018}. Moreover, TMD has much greater prevalence in females than in males and is believed to have some sex-specific pathophysiologic mechanisms~\citep{Bueno2018}. In this analysis, we wanted to explore the comparative performance of our method \texttt{pglmm} in selecting important sex-specific predictors of TMD and its performance predicting the risk of painful TMD in independent subjects from two replication cohorts, the OPPERA II Chronic TMD Replication case-control study, and the Complex Persistent Pain Conditions (CPPC): Unique and Shared Pathways of Vulnerability study, using the OPPERA cohort as discovery cohort. Sample sizes and distribution of sex, cases and ancestry for the three studies are shown in Table \ref{tab:TMD}, and further details on study design, recruitment, subject characteristics, and phenotyping for each study are provided in the Supplementary
Materials of \cite{Smith2018} (available at \url{http://links.lww.com/PAIN/A688}).

\begin{table}[!h]
\centering
\caption{Demographic data for the OPPERA training cohort, and for the OPPERA2 and PPG test cohorts.}\label{tab:TMD}
\begin{tabular}{l||lll}
\hline
& \multicolumn{3}{c}{Study name} \\ \hline\hline
 & OPPERA & OPPERA2 & PPG \\ \hline
N (\% female)       & 3030 (64.6)   & 1342 (66.0)   & 390 (84.4)   \\ 
Cases (\%)          & 999 (33.0)     & 444 (33.0)    & 164 (42.0)    \\ 
Ancestry (\% white) & 61         & 79            & 68           \\ \hline
\end{tabular}
\end{table}

We used the imputed data described in \cite{Smith2018}. Genotypes were imputed to the 1000 Genomes Project phase 3 reference panel using the software packages SHAPEIT \citep{Delaneau2011} for prephasing and IMPUTE version 2 \citep{Howie2009}. For each cohort independently, we assessed imputation quality taking into account the number of minor alleles as well as the information score such that a SNP with rare MAF must pass a higher quality information threshold for inclusion. 
After merging all three cohorts, we filtered for Hardy-Weinberg equilibrium (HWE) separately in cases and controls, using a more strict threshold among cases to avoid discarding disease-associated SNPs that are possibly under selection~\citep{Marees2018} ($<10^{-6}$ in controls, $<10^{-11}$ in cases). We filtered using a SNP call rate greater than $95\%$ on the combined dataset to retain imputed variants present in all cohorts, which resulted in a total of $4.8M$ imputed SNPs. PCs and kinship matrices were computed using the merged genotype data, following the same steps as described for the imputed data. To reduce the number of candidate predictors in the regularized models, we performed a first screening by testing genome-wide association with TMD for subjects in the OPPERA discovery cohort using PLINK~\citep{Chang2015}. We fitted a logistic regression for additive SNP effects, with age, sex and enrollment site as covariates and the first 10 PCs to account for population stratification, and retained all SNPs with a p-value below 0.05, which resulted in a total of 243K predictors.

We present in Table \ref{tab:gwas_snps} the estimated odds ratios (OR) by each method, \texttt{pglmm}, \texttt{gesso} and \texttt{glmnet}, for the selected SNPs for both main and GEI effects. Of note, it was not possible to use the \texttt{glinternet} package due to computational considerations, its memory requirement being too large for the joint analysis of the 243K preselected predictors. All three methods selected the imputed insertion/deletion (indel) polymorphism on chromosome 4 at position 146,211,844 (rs5862730), which was the only reported SNP that reached genome-wide significance in the full OPPERA cohort ($OR = 1.4$, $95\%$ confidence interval (CI): $[1.26;1.61]$, $P =2.82\times10^{-8}$)~\citep{Smith2018}. In a females-only analysis, rs5862730 was likewise associated with TMD ($OR = 1.54$, $95\%$ CI: $[1.33;1.79]$, $P = 1.7\times10^{-8}$), and both \texttt{pglmm} and \texttt{gesso} selected the GEI term between rs5862730 and sex.

Moreover, we present in Table \ref{tab:TMD} the AUC in the training and test cohorts, the number of predictors selected in each model and the total computation time to fit each method. We see that \texttt{pglmm} has the highest AUC on the training data, as well as the best predictive performance on the PPG cohort alone. On the other hand, \texttt{glmnet} and \texttt{gesso} both have a greater predictive performance in the OPPERA2 cohort compared to \texttt{pglmm}. When combining the predictions for OPPERA2 and PPG cohorts, all three methods have similar predictive performance. In term of the number of predictors selected by each model, \texttt{glmnet} has selected two SNPs with important main effects and no GEI effects, while \texttt{gesso} has selected the highest number of predictors, that is a total of 13 SNPs with both main and GEI effects. On the other hand, our proposed method \texttt{pglmm} has selected a total of 7 SNPs, among which 3 had a selected GEI effect with sex. Finally, we report for each method the computationnal time to fit the model on the training cohort using 10-folds cross-validation. While \texttt{glmnet} only took two hours to fit, it failed to retrieve any potentially important GEI effects between TMD and sex, albeit we note that it had a similar predictive performance than the hierarchical methods on the combined test sets. On the other hand, \texttt{pglmm} had the highest computational time required to fit the model, because it requires iteratively estimating a random effects vector of size $n=3030$, while both \texttt{glmnet} and \texttt{gesso} only require to estimate a vector of fixed effects of size 10 for the PCs. However, \texttt{pglmm} had the highest AUC on the train set, and was able to retrieve potentially important GEI effects for some of the select SNPs in the model, while selecting half as many predictors than \texttt{gesso}.

\begin{table}[ht]
\centering
\caption{Selected SNPs by each method with their estimated odds ratios (OR) for the main effects ($\beta$) and GEI effects ($\gamma$) from the TMD real data analysis.}\label{tab:gwas_snps}
\begin{tabular}{llccccccccc}
  \hline
&  & \multicolumn{3}{c}{\texttt{pglmm}} & \multicolumn{3}{c}{\texttt{gesso}} & \multicolumn{1}{c}{\texttt{glmnet}} \\ 
Chromosome & Position & OR$_\beta$ & OR$_\gamma$ & OR$_{\beta+\gamma}$ & OR$_\beta$ & OR$_\gamma$ & OR$_{\beta+\gamma}$ & OR$_\beta$\\ 
  \hline
    3 & 5,046,726 & - & - & - & 1.0042 & 1.0087 & 1.0129 & - \\ 
      3 & 153,536,154 & 1.0020 & - & - & - & - & - & - \\ 
      4 & 42,549,777 & 1.0068 & 1.0042 & 1.0110 & 1.0029 & 1.0060 & 1.0089 & - \\ 
      \textbf{4} & \textbf{146,211,844} & 1.0252 & 1.0448 & 1.0712 & 1.0261 & 1.0553 & 1.0829 & 1.0312 \\ 
     11 & 17,086,381 & 1.0076 & - & - & 1.0014 & 1.0029 & 1.0042 & - \\ 
     11 & 132,309,606 & 0.9965 & - & - & - & - & - & - \\ 
     12 & 19,770,625 & - & - & - & 1.0045 & 1.0094 & 1.0140 & - \\ 
     12 & 47,866,802 & 1.0184 & 1.0001 & 1.0184 & - & - & - & 1.0140 \\ 
     12 & 47,870,741 & - & - & - & 1.0152 & 1.0320 & 1.0477 & - \\ 
     14 & 24,345,235 & 1.0013 & - & - & - & - & - & -\\ 
     16 & 81,155,867 & - & - & - & 1.0039 & 1.0082 & 1.0122 & - \\ 
     17 & 46,592,346 & - & - & - & 1.0025 & 1.0052 & 1.0077 & - \\ 
     17 & 52,888,414 & - & - & - & 1.0005 & 1.0011 & 1.0017 & - \\ 
     17 & 69,061,947 & - & - & - & 1.0021 & 1.0043 & 1.0064 & - \\ 
     18 & 36,210,549 & - & - & - & 1.0186 & 1.0392 & 1.0585 & - \\ 
     19 & 37,070,882 & - & - & - & 1.0020 & 1.0042 & 1.0062 & - \\ 
     21 & 32,760,615 & - & - & - & 1.0051 & 1.0107 & 1.0159 & - \\ 
   \hline
\end{tabular}
\end{table}

\begin{table}[]
\centering
\caption{Area under the roc curve (AUC), model size and computational time for the analysis of TMD.}\label{tab:TMDres}
\begin{tabular}{l||cccccccc}
\hline
  &  AUC$_{train}$ &  \multicolumn{3}{c}{AUC$_{test}$} & \multicolumn{2}{c}{
 Model size} & Computational \\ \cline{3-5}
 Method &  OPPERA &  OPPERA2 & PPG & OPPERA2+PPG & Main effects & GEI effects &  time (hours) \\ 
 \hline
 \texttt{glmnet} & 0.722 & 0.587 & 0.632 & 0.551 & 2 & 0 & 2\\ 
\texttt{gesso} & 0.725 & 0.586 & 0.630 & 0.551 & 13 & 13 & 9\\ 
\texttt{pglmm} & 0.867 & 0.512 & 0.652 & 0.550 & 9 & 5 & 47\\
\hline

\end{tabular}
\end{table}

\clearpage
\section{Discussion}
We have developed a unified approach based on regularized PQL estimation, for selecting important predictors and GEI effects in high-dimensional GWAS data, accounting for population structure, close relatedness, shared environmental exposure and binary nature of the trait. We proposed to combine PQL estimation with a CAP for hierarchical selection of main genetic and GEI effects, and derived a proximal Newton-type algorithm with block coordinate descent to find coordinate-wise updates. We showed that for all simulation scenarios, \pglmm~always selected the lowest number of predictors in the model, had the lowest FPR and FDR, and had the highest prediction accuracy in independent subjects as measured by the AUC. Thus, compared to other penalized methods, \pglmm~enforced sparsity by controlling the number of false positives in the model while having the best predictive performance. Moreover, using real data from the OPPERA study to explore the comparative performance of our method in selecting important predictors of TMD, we found that our method was able to retrieve a previously reported significant loci in a combined or sex-segregated GWAS.

A limitation of \texttt{pglmm} compared to a logistic lasso or group lasso with PC adjustment is the computational cost of performing multiple matrix calculations that comes from incorporating a GSM to account for population structure and relatedness between individuals. 
These computations become prohibitive when the sample size increases, and this may hinder the use of random effects in hierachichal selection of both genetic and GEI fixed effects in genetic association studies. Solutions to explore in order to increase computation speed and decrease memory usage would be the use of conjugate gradient methods with a diagonal preconditioner matrix, as proposed by~\citet{zhouEfficientlyControllingCasecontrol2018}, and the use of sparse GSMs to adjust for the sample relatedness~\citep{jiang2019}.

In this study, we focused solely on the sparse group lasso as a hierarchical regularization penalty. However, it is known that estimated effects by lasso will have large biases because the resulting shrinkage is constant irrespective of the magnitude of the effects. Alternative regularizations like the Smoothly Clipped Absolute Deviation (SCAD)~\citep{Fan2001} and Minimax Concave Penalty (MCP)~\citep{MCP} could be explored, although we note that both SCAD and MCP require tuning an additional parameter which controls the relaxation rate of the penalty. Another alternative includes refitting the sparse group lasso penalty on the active set of predictors only, similarly to the relaxed lasso, which has shown to produce sparser models with equal or lower prediction loss than the regular lasso estimator for high-dimensional data~\citep{Meinshausen2007}. Finally, it would also be of interest to explore if joint selection of fixed and random effects could result in better selection and/or predictive performance, as including high-dimensional random effects can potentially lead to a near singular covariance matrix. Future work includes tuning the generalized ridge regularization on the random effects~\citep{Shen2013}, or replacing it by a lasso regularization to perform selection of individual random effects~\citep{huiJointSelectionMixed2017, Bondell2010}.


\section*{Acknowledgments}
This work was supported by the Fonds de recherche Québec-Santé [267074 to K.O.]; and the Natural Sciences and Engineering Research Council of Canada [RGPIN-2019-06727 to K.O., RGPIN-2020-05133 to S.B.].

This study was enabled in part by support provided by Calcul Québec (\url{https://www.calculquebec.ca}) and Compute Canada (\url{https://www.computecanada.ca}). The authors would like to recognize the contribution from S.B. Smith, L. Diatchenko and the analytical team at McGill University, in particular M. Parisien, for providing support with the data from OPPERA, OPPERA II and CPPC studies. OPPERA was supported by the National Institute of Dental and Craniofacial Research (NIDCR; https://www.nidcr.nih.gov/): grant number U01DE017018. The OPPERA program also acknowledges resources specifically provided for this project by the respective host universities: University at Buffalo, University of Florida, University of Maryland–Baltimore, and University of North Carolina–Chapel Hill. Funding for genotyping was provided by NIDCR through a contract to the Center for Inherited Disease
Research at Johns Hopkins University (HHSN268201200008I). Data from the OPPERA study are available through the NIH dbGaP: phs000796.v1.p1 and phs000761.v1.p1. L. Diatchenko and the analytical team at McGill University were supported by the Canadian Excellence Research Chairs (CERC)
Program grant\\ (http://www.cerc.gc.ca/home-accueil-eng.aspx,
CERC09). The Complex Persistent Pain Conditions: Unique and Shared
Pathways of Vulnerability Program Project were supported by
NIH/National Institute of Neurological Disorders and Stroke
(NINDS; https://www.ninds.nih.gov) grant NS045685 to the
University of North Carolina at Chapel Hill, and genotyping was funded by the Canadian Excellence Research Chairs (CERC)
Program (grant CERC09).
The OPPERA II study was supported by the NIDCR under Award
Number U01DE017018, and genotyping was funded by the
Canadian Excellence Research Chairs (CERC) Program (grant
CERC09).



\section*{Data availability statement}\label{sec:dataavail}
Our Julia package \texttt{PenalizedGLMM} and codes for simulating data are available on github \url{https://github.com/julstpierre/PenalizedGLMM}.

\bibliographystyle{natbib}
\bibliography{biblio}

\begin{thebibliography}{}

\bibitem[Bhatnagar {\em et~al.}(2020)Bhatnagar, Yang, Lu, Schurr, Loredo-Osti,
  Forest, Oualkacha, and Greenwood]{Bhatnagar2020}
Bhatnagar, S.~R.  {\em et~al.} (2020).
\newblock Simultaneous {SNP} selection and adjustment for population structure
  in high dimensional prediction models.
\newblock {\em {PLOS} Genetics\/}, {\bf 16}(5), e1008766.

\bibitem[Bien {\em et~al.}(2013)Bien, Taylor, and Tibshirani]{Bien2013}
Bien, J.  {\em et~al.} (2013).
\newblock A lasso for hierarchical interactions.
\newblock {\em The Annals of Statistics\/}, {\bf 41}(3).

\bibitem[Bondell {\em et~al.}(2010)Bondell, Krishna, and Ghosh]{Bondell2010}
Bondell, H.~D.  {\em et~al.} (2010).
\newblock Joint variable selection for fixed and random effects in linear
  mixed-effects models.
\newblock {\em Biometrics\/}, {\bf 66}(4), 1069--1077.

\bibitem[Breslow and Clayton(1993)Breslow and
  Clayton]{breslowApproximateInferenceGeneralized1993}
Breslow, N.~E. and Clayton, D.~G. (1993).
\newblock Approximate {{Inference}} in {{Generalized Linear Mixed Models}}.
\newblock {\em Journal of the American Statistical Association\/}, {\bf
  88}(421), 9--25.

\bibitem[Bueno {\em et~al.}(2018)Bueno, Pereira, Pattussi, Grossi, and
  Grossi]{Bueno2018}
Bueno, C.~H.  {\em et~al.} (2018).
\newblock Gender differences in temporomandibular disorders in adult
  populational studies: A systematic review and meta-analysis.
\newblock {\em Journal of Oral Rehabilitation\/}, {\bf 45}(9), 720--729.

\bibitem[Chang {\em et~al.}(2015)Chang, Chow, Tellier, Vattikuti, Purcell, and
  Lee]{Chang2015}
Chang, C.~C.  {\em et~al.} (2015).
\newblock Second-generation {PLINK}: rising to the challenge of larger and
  richer datasets.
\newblock {\em {GigaScience}\/}, {\bf 4}(1).

\bibitem[Chen {\em et~al.}(2016)Chen, Wang, Conomos, Stilp, Li, Sofer, Szpiro,
  Chen, Brehm, Celed{\'{o}}n, Redline, Papanicolaou, Thornton, Laurie, Rice,
  and Lin]{Chen2016}
Chen, H.  {\em et~al.} (2016).
\newblock Control for population structure and relatedness for binary traits in
  genetic association studies via logistic mixed models.
\newblock {\em The American Journal of Human Genetics\/}, {\bf 98}(4),
  653--666.

\bibitem[Cox(1984)Cox]{Cox1984}
Cox, D.~R. (1984).
\newblock Interaction.
\newblock {\em International Statistical Review / Revue Internationale de
  Statistique\/}, {\bf 52}(1), 1.

\bibitem[Delaneau {\em et~al.}(2011)Delaneau, Marchini, and
  Zagury]{Delaneau2011}
Delaneau, O.  {\em et~al.} (2011).
\newblock A linear complexity phasing method for thousands of genomes.
\newblock {\em Nature Methods\/}, {\bf 9}(2), 179--181.

\bibitem[Dudbridge and Fletcher(2014)Dudbridge and Fletcher]{Dudbridge2014}
Dudbridge, F. and Fletcher, O. (2014).
\newblock Gene-environment dependence creates spurious gene-environment
  interaction.
\newblock {\em The American Journal of Human Genetics\/}, {\bf 95}(3),
  301--307.

\bibitem[Fan and Li(2001)Fan and Li]{Fan2001}
Fan, J. and Li, R. (2001).
\newblock Variable selection via nonconcave penalized likelihood and its oracle
  properties.
\newblock {\em Journal of the American Statistical Association\/}, {\bf
  96}(456), 1348--1360.

\bibitem[Fang {\em et~al.}(2023)Fang, Li, Zhang, Xu, and Ma]{Fang2023}
Fang, K.  {\em et~al.} (2023).
\newblock Pathological imaging-assisted cancer gene{\textendash}environment
  interaction analysis.
\newblock {\em Biometrics\/}.

\bibitem[Friedman {\em et~al.}(2010)Friedman, Hastie, and Tibshirani]{glmnet}
Friedman, J.  {\em et~al.} (2010).
\newblock Regularization paths for generalized linear models via coordinate
  descent.
\newblock {\em Journal of Statistical Software\/}, {\bf 33}(1), 1--22.

\bibitem[Gilmour {\em et~al.}(1995)Gilmour, Thompson, and Cullis]{Gilmour1995}
Gilmour, A.~R.  {\em et~al.} (1995).
\newblock Average information {REML}: An efficient algorithm for variance
  parameter estimation in linear mixed models.
\newblock {\em Biometrics\/}, {\bf 51}(4), 1440.

\bibitem[Hoffman(2013)Hoffman]{Hoffman2013}
Hoffman, G.~E. (2013).
\newblock Correcting for population structure and kinship using the linear
  mixed model: Theory and extensions.
\newblock {\em {PLoS} {ONE}\/}, {\bf 8}(10), e75707.

\bibitem[Howie {\em et~al.}(2009)Howie, Donnelly, and Marchini]{Howie2009}
Howie, B.~N.  {\em et~al.} (2009).
\newblock A flexible and accurate genotype imputation method for the next
  generation of genome-wide association studies.
\newblock {\em {PLoS} Genetics\/}, {\bf 5}(6), e1000529.

\bibitem[Hu {\em et~al.}(2019)Hu, Lu, Zhou, and Zhou]{Hu2019}
Hu, L.  {\em et~al.} (2019).
\newblock {MM} {ALGORITHMS} {FOR} {VARIANCE} {COMPONENT} {ESTIMATION} {AND}
  {SELECTION} {IN} {LOGISTIC} {LINEAR} {MIXED} {MODEL}.
\newblock {\em Statistica Sinica\/}.

\bibitem[Hui {\em et~al.}(2017)Hui, M{\"u}ller, and
  Welsh]{huiJointSelectionMixed2017}
Hui, F. K.~C.  {\em et~al.} (2017).
\newblock Joint {{Selection}} in {{Mixed Models}} using {{Regularized PQL}}.
\newblock {\em Journal of the American Statistical Association\/}, {\bf
  112}(519), 1323--1333.

\bibitem[Jiang {\em et~al.}(2019)Jiang, Zheng, Qi, Kemper, Wray, Visscher, and
  Yang]{jiang2019}
Jiang, L.  {\em et~al.} (2019).
\newblock A resource-efficient tool for mixed model association analysis of
  large-scale data.
\newblock {\em Nature genetics\/}, {\bf 51}(12), 1749--1755.

\bibitem[Koenig {\em et~al.}(2023)Koenig, Yohannes, Nkambule, Goodrich, Kim,
  Zhao, Wilson, Tiao, Hao, Sahakian, Chao, Talkowski, Daly, Brand, Karczewski,
  Atkinson, and and]{Koenig2023}
Koenig, Z.  {\em et~al.} (2023).
\newblock A harmonized public resource of deeply sequenced diverse human
  genomes.

\bibitem[Kooij(2007)Kooij]{Kooij}
Kooij, A. (2007).
\newblock Prediction accuracy and stability of regression with optimal scaling
  transformations.
\newblock {\em PhD thesis. Faculty of Social and Behavioural Sciences, Leiden
  University\/}.

\bibitem[Liang {\em et~al.}(2022)Liang, Cohen, Heinsfeld, Pestilli, and
  McDonald]{liang2022sparsegl}
Liang, X.  {\em et~al.} (2022).
\newblock sparsegl: An r package for estimating sparse group lasso.
\newblock {\em arXiv preprint arXiv:2208.02942\/}.

\bibitem[Lim and Hastie(2015)Lim and Hastie]{Lim2015}
Lim, M. and Hastie, T. (2015).
\newblock Learning interactions via hierarchical group-lasso regularization.
\newblock {\em Journal of Computational and Graphical Statistics\/}, {\bf
  24}(3), 627--654.

\bibitem[Maixner {\em et~al.}(2011)Maixner, Diatchenko, Dubner, Fillingim,
  Greenspan, Knott, Ohrbach, Weir, and Slade]{Maixner2011}
Maixner, W.  {\em et~al.} (2011).
\newblock Orofacial pain prospective evaluation and risk assessment study
  {\textendash} the {OPPERA} study.
\newblock {\em The Journal of Pain\/}, {\bf 12}(11), T4--T11.e2.

\bibitem[Marees {\em et~al.}(2018)Marees, de~Kluiver, Stringer, Vorspan, Curis,
  Marie-Claire, and Derks]{Marees2018}
Marees, A.~T.  {\em et~al.} (2018).
\newblock A tutorial on conducting genome-wide association studies: Quality
  control and statistical analysis.
\newblock {\em International Journal of Methods in Psychiatric Research\/},
  {\bf 27}(2), e1608.

\bibitem[Meinshausen(2007)Meinshausen]{Meinshausen2007}
Meinshausen, N. (2007).
\newblock Relaxed lasso.
\newblock {\em Computational Statistics \& Data Analysis\/}, {\bf 52}(1),
  374--393.

\bibitem[Mukherjee {\em et~al.}(2009)Mukherjee, Ahn, Gruber, Ghosh, and
  Chatterjee]{Mukherjee2009}
Mukherjee, B.  {\em et~al.} (2009).
\newblock Case-control studies of gene-environment interaction: Bayesian design
  and analysis.
\newblock {\em Biometrics\/}, {\bf 66}(3), 934--948.

\bibitem[{\O}deg{\aa}rd {\em et~al.}(2018){\O}deg{\aa}rd, Indahl,
  Strand{\'{e}}n, and Meuwissen]{degrd2018}
{\O}deg{\aa}rd, J.  {\em et~al.} (2018).
\newblock Large-scale genomic prediction using singular value decomposition of
  the genotype matrix.
\newblock {\em Genetics Selection Evolution\/}, {\bf 50}(1).

\bibitem[Price {\em et~al.}(2010)Price, Zaitlen, Reich, and
  Patterson]{Price2010}
Price, A.~L.  {\em et~al.} (2010).
\newblock New approaches to population stratification in genome-wide
  association studies.
\newblock {\em Nature Reviews Genetics\/}, {\bf 11}(7), 459--463.

\bibitem[Shen {\em et~al.}(2013)Shen, Alam, Fikse, and
  R\"{o}nneg{\aa}rd]{Shen2013}
Shen, X.  {\em et~al.} (2013).
\newblock A novel generalized ridge regression method for quantitative
  genetics.
\newblock {\em Genetics\/}, {\bf 193}(4), 1255--1268.

\bibitem[Shi {\em et~al.}(2020)Shi, O'Brien, and Weinberg]{Shi2020}
Shi, M.  {\em et~al.} (2020).
\newblock Interactions between a polygenic risk score and non-genetic risk
  factors in young-onset breast cancer.
\newblock {\em Scientific Reports\/}, {\bf 10}(1).

\bibitem[Smith {\em et~al.}(2018)Smith, Parisien, Bair, Belfer,
  Chabot-Dor{\'{e}}, Gris, Khoury, Tansley, Torosyan, Zaykin, Bernhardt,
  de~Oliveira~Serrano, Gracely, Jain, J\"{a}rvelin, Kaste, Kerr, Kocher,
  L\"{a}hdesm\"{a}ki, Laniado, Laurie, Laurie, M\"{a}nnikk\"{o}, Meloto,
  Nackley, Nelson, Pesonen, Ribeiro-Dasilva, Rizzatti-Barbosa, Sanders,
  Schwahn, Sipil\"{a}, Sofer, Teumer, Mogil, Fillingim, Greenspan, Ohrbach,
  Slade, Maixner, and Diatchenko]{Smith2018}
Smith, S.~B.  {\em et~al.} (2018).
\newblock Genome-wide association reveals contribution of {MRAS} to painful
  temporomandibular disorder in males.
\newblock {\em Pain\/}, {\bf 160}(3), 579--591.

\bibitem[St-Pierre {\em et~al.}(2023)St-Pierre, Oualkacha, and
  Bhatnagar]{StPierre2023}
St-Pierre, J.  {\em et~al.} (2023).
\newblock Efficient penalized generalized linear mixed models for variable
  selection and genetic risk prediction in high-dimensional data.
\newblock {\em Bioinformatics\/}, {\bf 39}(2).

\bibitem[Sul {\em et~al.}(2016)Sul, Bilow, Yang, Kostem, Furlotte, He, and
  Eskin]{sul2016}
Sul, J.~H.  {\em et~al.} (2016).
\newblock Accounting for {{Population Structure}} in {{Gene}}-by-{{Environment
  Interactions}} in {{Genome}}-{{Wide Association Studies Using Mixed Models}}.
\newblock {\em PLOS Genetics\/}, {\bf 12}(3), e1005849.

\bibitem[Tibshirani {\em et~al.}(2011)Tibshirani, Bien, Friedman, Hastie,
  Simon, Taylor, and Tibshirani]{Tibshirani2011}
Tibshirani, R.  {\em et~al.} (2011).
\newblock Strong rules for discarding predictors in lasso-type problems.
\newblock {\em Journal of the Royal Statistical Society: Series B (Statistical
  Methodology)\/}, {\bf 74}(2), 245--266.

\bibitem[Wu and Lange(2008)Wu and Lange]{Wu2008}
Wu, T.~T. and Lange, K. (2008).
\newblock Coordinate descent algorithms for lasso penalized regression.
\newblock {\em The Annals of Applied Statistics\/}, {\bf 2}(1).

\bibitem[Yu {\em et~al.}(2005)Yu, Pressoir, Briggs, Bi, Yamasaki, Doebley,
  McMullen, Gaut, Nielsen, Holland, Kresovich, and Buckler]{Yu2005}
Yu, J.  {\em et~al.} (2005).
\newblock A unified mixed-model method for association mapping that accounts
  for multiple levels of relatedness.
\newblock {\em Nature Genetics\/}, {\bf 38}(2), 203--208.

\bibitem[Zemlianskaia {\em et~al.}(2022)Zemlianskaia, Gauderman, and
  Lewinger]{Zemlianskaia2022}
Zemlianskaia, N.  {\em et~al.} (2022).
\newblock A scalable hierarchical lasso for gene{\textendash}environment
  interactions.
\newblock {\em Journal of Computational and Graphical Statistics\/}, pages
  1--13.

\bibitem[Zhang(2010)Zhang]{MCP}
Zhang, C.-H. (2010).
\newblock {Nearly unbiased variable selection under minimax concave penalty}.
\newblock {\em The Annals of Statistics\/}, {\bf 38}(2), 894 -- 942.

\bibitem[Zhao {\em et~al.}(2009)Zhao, Rocha, and Yu]{Zhao2009}
Zhao, P.  {\em et~al.} (2009).
\newblock The composite absolute penalties family for grouped and hierarchical
  variable selection.
\newblock {\em The Annals of Statistics\/}, {\bf 37}(6A).

\bibitem[Zhou {\em et~al.}(2019)Zhou, Hu, Zhou, and Lange]{Zhou2019}
Zhou, H.  {\em et~al.} (2019).
\newblock {MM} algorithms for variance components models.
\newblock {\em Journal of Computational and Graphical Statistics\/}, {\bf
  28}(2), 350--361.

\bibitem[Zhou {\em et~al.}(2018)Zhou, Nielsen, Fritsche, Dey, Gabrielsen,
  Wolford, LeFaive, VandeHaar, Gagliano, Gifford, Bastarache, Wei, Denny, Lin,
  Hveem, Kang, Abecasis, Willer, and
  Lee]{zhouEfficientlyControllingCasecontrol2018}
Zhou, W.  {\em et~al.} (2018).
\newblock Efficiently controlling for case-control imbalance and sample
  relatedness in large-scale genetic association studies.
\newblock {\em Nature Genetics\/}, {\bf 50}(9), 1335--1341.

\end{thebibliography}


\begin{thebibliography}{7}
\providecommand{\natexlab}[1]{#1}
\providecommand{\url}[1]{\texttt{#1}}
\expandafter\ifx\csname urlstyle\endcsname\relax
  \providecommand{\doi}[1]{doi: #1}\else
  \providecommand{\doi}{doi: \begingroup \urlstyle{rm}\Url}\fi

\bibitem[{\O}deg{\aa}rd et~al.(2018){\O}deg{\aa}rd, Indahl, Strand{\'{e}}n, and
  Meuwissen]{degrd2018}
J{\o}rgen {\O}deg{\aa}rd, Ulf Indahl, Ismo Strand{\'{e}}n, and Theo H.~E.
  Meuwissen.
\newblock Large-scale genomic prediction using singular value decomposition of
  the genotype matrix.
\newblock \emph{Genetics Selection Evolution}, 50\penalty0 (1), February 2018.
\newblock \doi{10.1186/s12711-018-0373-2}.
\newblock URL \url{https://doi.org/10.1186/s12711-018-0373-2}.

\bibitem[Kooij(2007)]{Kooij}
Anita Kooij.
\newblock Prediction accuracy and stability of regression with optimal scaling
  transformations.
\newblock \emph{PhD thesis. Faculty of Social and Behavioural Sciences, Leiden
  University}, 01 2007.

\bibitem[Friedman et~al.(2010)Friedman, Hastie, and Tibshirani]{glmnet}
Jerome Friedman, Trevor Hastie, and Robert Tibshirani.
\newblock Regularization paths for generalized linear models via coordinate
  descent.
\newblock \emph{Journal of Statistical Software}, 33\penalty0 (1):\penalty0
  1--22, 2010.
\newblock URL \url{https://www.jstatsoft.org/v33/i01/}.

\bibitem[Tibshirani et~al.(2011)Tibshirani, Bien, Friedman, Hastie, Simon,
  Taylor, and Tibshirani]{Tibshirani2011}
Robert Tibshirani, Jacob Bien, Jerome Friedman, Trevor Hastie, Noah Simon,
  Jonathan Taylor, and Ryan~J. Tibshirani.
\newblock Strong rules for discarding predictors in lasso-type problems.
\newblock \emph{Journal of the Royal Statistical Society: Series B (Statistical
  Methodology)}, 74\penalty0 (2):\penalty0 245--266, November 2011.
\newblock \doi{10.1111/j.1467-9868.2011.01004.x}.
\newblock URL \url{https://doi.org/10.1111/j.1467-9868.2011.01004.x}.

\bibitem[Liang et~al.(2022)Liang, Cohen, Heinsfeld, Pestilli, and
  McDonald]{liang2022sparsegl}
Xiaoxuan Liang, Aaron Cohen, Anibal~Sol{\'o}n Heinsfeld, Franco Pestilli, and
  Daniel~J McDonald.
\newblock sparsegl: An r package for estimating sparse group lasso.
\newblock \emph{arXiv preprint arXiv:2208.02942}, 2022.

\bibitem[Bhatnagar et~al.(2020)Bhatnagar, Yang, Lu, Schurr, Loredo-Osti,
  Forest, Oualkacha, and Greenwood]{Bhatnagar2020}
Sahir~R. Bhatnagar, Yi~Yang, Tianyuan Lu, Erwin Schurr, JC~Loredo-Osti, Marie
  Forest, Karim Oualkacha, and Celia M.~T. Greenwood.
\newblock Simultaneous {SNP} selection and adjustment for population structure
  in high dimensional prediction models.
\newblock \emph{{PLOS} Genetics}, 16\penalty0 (5):\penalty0 e1008766, May 2020.
\newblock \doi{10.1371/journal.pgen.1008766}.
\newblock URL \url{https://doi.org/10.1371/journal.pgen.1008766}.

\bibitem[Wu and Lange(2008)]{Wu2008}
Tong~Tong Wu and Kenneth Lange.
\newblock Coordinate descent algorithms for lasso penalized regression.
\newblock \emph{The Annals of Applied Statistics}, 2\penalty0 (1), March 2008.
\newblock \doi{10.1214/07-aoas147}.
\newblock URL \url{https://doi.org/10.1214/07-aoas147}.

\end{thebibliography}

\clearpage

\begin{appendix}
\section{Algorithmic methods}\label{appendix:1}
\subsection{Updates for \texorpdfstring{$\bm{\tilde{\delta}}$}{delta}}
The gradient and Hessian of $f(\bm{\Theta};\bm{{\delta}})$ are given by 
\begin{align*}
\nabla_{\bm{\delta}} f(\bm{\Theta};\bm{{\delta}}) &= -\hat\phi^{-1}\bm{U}^\intercal(\bm y - \bm\mu) + \bm{\Lambda}^{-1}\bm{\delta}, \\
\nabla^2_{\bm{\delta}} f(\bm{\Theta};\bm{{\delta}}) &= \hat\phi^{-1}\bm{U}^\intercal\bm{\Delta}^{-1}\bm{U} + \bm{\Lambda}^{-1}.
\end{align*}
This leads to the Newton updates
\begin{align}\label{eq:Newton}
\tilde{\bm{\delta}}^{(t+1)} &= \tilde{\bm{\delta}}^{(t)} - [\nabla^2_{\bm{\delta}} f(\bm{\Theta}|\tilde{\bm{\delta}}^{(t)})]^{-1} \nabla_{\bm{\delta}} f(\bm{\Theta}|\tilde{\bm{\delta}}^{(t)}) \nonumber \\
&= \tilde{\bm{\delta}}^{(t)} + \left[\hat\phi^{-1}\bm{U}^\intercal\bm{\Delta}^{-(t)}\bm{U} + \bm{\Lambda}^{-1}\right]^{-1}\left(\hat\phi^{-1}\bm{U}^\intercal(\bm y - \bm\mu^{(t)}) - \bm{\Lambda}^{-1}\tilde{\bm{\delta}}^{(t)}\right) \nonumber  \\
&= \left[\bm{U}^\intercal\bm{\Delta}^{-(t)}\bm{U} + \hat\phi\bm{\Lambda}^{-1}\right]^{-1}\bm{U}^\intercal\bm{\Delta}^{-(t)}\left(\bm{\Delta}^{(t)}(\bm y - \bm\mu^{(t)}) + \bm{U}\tilde{\bm{\delta}}^{(t)}\right),
\end{align}
which requires repeatedly inverting the $n\times n$ matrix $\bm{\Sigma}^{(t)} := \bm{U}^\intercal\bm{\Delta}^{-(t)}\bm{U} + \hat\phi\bm{\Lambda}^{-1}$ with complexity $O(n^3)$ where $n$ is the sample size. 
Defining the working vector $\tilde{\bm{Y}} = \bm{X}\bm{\Theta}^{(t)} + \bm{U}\tilde{\bm{\delta}}^{(t)} + \bm{\Delta}^{(t)}(\bm y - \bm\mu^{(t)})$, where $\bm{X}\bm{\Theta}=\bm{Z} \bm\theta + \bm D \alpha + \bm{G} \bm\beta + (\bm D \odot \bm{G}) \bm{\gamma}$, the Newton updates in \eqref{eq:Newton} can be rewritten as
\begin{align*}
\tilde{\bm{\delta}}^{(t+1)} &= \left[\bm{U}^\intercal\bm{\Delta}^{-(t)}\bm{U} + \hat\phi\bm{\Lambda}^{-1}\right]^{-1}\bm{U}^\intercal\bm{\Delta}^{-(t)}\left(\tilde{\bm{Y}} - \bm{X} \bm\Theta^{(t)}\right),
\end{align*}
which can be equivalently obtained as the solutions to the following generalized ridge weighted least-squares (WLS) problem
\begin{align}\label{eq:irwls}
\tilde{\bm{\delta}}^{(t+1)} &=  \underset{\bm \delta}{\textrm{argmin }} \hat{\phi}^{-1}\left(\tilde{\bm{Y}} - \bm{X}\bm{\Theta}^{(t)} - \bm{U\delta}\right)^\intercal \bm{\Delta}^{-(t)} \left(\tilde{\bm{Y}} - \bm{X}\bm{\Theta}^{(t)} - \bm{U\delta} \right) + \bm{\delta}^\intercal\bm{\Lambda}^{-1}\bm \delta.
\end{align}
Equation \eqref{eq:irwls} is analogous to the principal component ridge regression (PCRR) model~\citep{degrd2018}, and demonstrates that PCA and MMs indeed share the same underlying model. At last, to solve \eqref{eq:irwls} without repeatedly inverting the $n\times n$ matrix $\bm{\Sigma}^{(t)} := \bm{U}^\intercal\bm{\Delta}^{-(t)}\bm{U} + \hat\phi\bm{\Lambda}^{-1}$, we propose using a coordinate descent algorithm~\citep{Kooij}, for which each coordinate's updates are given, for $j=1,...,n$, by
\begin{align}\label{eq:coord_desc}
\tilde{\delta_j} \leftarrow \frac{\sum_{i=1}^n w_iU_{ij}\left(\tilde{Y}_i- \bm{X}_i\bm{\Theta}^{(t)} - \sum_{l\ne j}U_{il}\tilde{\delta_l}\right)}{\sum_{i=1}^n w_i U_{ij}^{2} + \hat{\phi}\Lambda_j^{-1}},   
\end{align}
where $w_i = \bm{\Delta}_{ii}^{-(t)}$. 

\subsection{Updates for \texorpdfstring{$\bm{\Theta}$}{Theta}}
Since the objective function in \eqref{eq:objfunc2} consists of a smooth convex function $f(\bm{\Theta};\bm{\delta})$ and a non-smooth convex regularizer $g(\bm{\Theta})$, we propose a proximal Newton algorithm with cyclic coordinate descent to find PQL regularized estimates for $\bm{\Theta}$, in the spirit of the proposed algorithm by~\citet{glmnet} for estimation of generalized linear models with convex penalties. Let again $\bm{X}\bm{\Theta}=\bm{Z} \bm\theta + \bm D \alpha + \bm{G} \bm\beta + (\bm D \odot \bm{G}) \bm{\gamma}$ and $\bm{{\Theta}}^{(t)}$ be the current iterate, the iterative step reduces to
\begin{align*}
\bm{{\Theta}}^{(t+1)} &= \underset{\bm{{\Theta}}}{\textrm{argmin }} \left\{\frac{1}{2s_t} \left\|\bm{{\Theta}}-\left(\bm{{\Theta}}^{(t)} - s_t \left[\nabla^2_{\bm \Theta} f(\bm{{\Theta}}^{(t)}|\bm{\tilde{\delta}})\right]^{-1}\nabla_{\bm \Theta} f(\bm{{\Theta}}^{(t)}|\bm{\tilde{\delta}})\right)\right\|_2^2 + g(\bm{\Theta})\right\} \\
&= \underset{\bm{{\Theta}}}{\textrm{argmin }} \left\{\frac{1}{2s_t} \left\|\bm{{\Theta}}- \left[\bm{X}^\intercal\bm{\Delta}^{-(t)}\bm{X}\right]^{-1}\bm{X}^\intercal\bm{\Delta}^{-(t)}\left(\bm{X\Theta}^{(t)} + s_t\bm{\Delta}^{(t)}(\bm y - \bm\mu^{(t)})\right)\right\|_2^2 + g(\bm{\Theta})\right\},
\end{align*}
where $s_t$ is a suitable step size. Defining the working vector $\tilde{\bm{Y}} = \bm{X}\bm{\Theta}^{(t)} + \bm{U}\tilde{\bm{\delta}}^{(t+1)} + s_t\bm{\Delta}^{(t)}(\bm y - \bm\mu^{(t)})$, we can again rewrite the minimization problem as a WLS problem where
\begin{align}\label{eq:proxnewton}
\bm{{\Theta}}^{(t+1)} &= \underset{\bm{{\Theta}}}{\textrm{argmin }} \left\{\frac{1}{2s_t} \left\|\bm{{\Theta}}- \left[\bm{X}^\intercal\bm{\Delta}^{-(t)}\bm{X}\right]^{-1}\bm{X}^\intercal\bm{\Delta}^{-(t)}\left(\bm{\tilde{Y}} - \bm{U}\tilde{\bm{\delta}}^{(t+1)}\right)\right\|_2^2 + g(\bm{\Theta})\right\} \nonumber \\
&= \underset{\bm{{\Theta}}}{\textrm{argmin }} \left\{\frac{1}{2s_t} \sum_{i=1}^n w_i\left(\tilde{Y}_i - \bm{X}_i\bm{\Theta} - \bm{U}_i\tilde{\bm{\delta}}^{(t+1)}\right)^2 + (1-\rho)\lambda\sum_j \|\beta_j, \gamma_j\|_2 + \rho\lambda\sum_j|\gamma_j|\right\},
\end{align}
where $w_i = \bm{\Delta}_{ii}^{-(t)}$. We use block coordinate descent and minimize \eqref{eq:proxnewton} with respect to each component of $\bm{\Theta}=\left(\bm{\theta}^\intercal, \alpha^\intercal, \bm\beta^\intercal ,\bm{\gamma}^\intercal\right)^\intercal$. In practice, we set $s_t=1$ and do not perform step-size optimization. We present in Appendix \ref{appendix:2} the detailed derivations and our block coordinate descent algorithm to obtain PQL regularized estimates for $\bm{\Theta}$.

\subsection{Strong rule}\label{subsec:strongrule}
In modern genome-wide studies, the number of genetic predictors is often very large, and assuming that most of the predictors effects are equal to 0, it would be desirable to discard them from the coordinate descent steps to speedup the optimization procedure.~\citet{Tibshirani2011} derived sequential strong rules that can be used when solving the lasso and lasso-type problems over a grid of tuning parameter values $\lambda_1 \ge \lambda_2 \ge \lambda_m$, and more details about the derivation of the sequential strong rule for the sparse group lasso can be found in~\citet{liang2022sparsegl}. Therefore, having already computed the solution $\hat{\bm{\Theta}}_{k-1}$ at $\lambda_{k-1}$, the sequential strong rule discards the $j^{th}$ genetic predictor from the optimization problem at $\lambda_k$ if 
\begin{align*}
\sqrt{\left(G_j^\intercal(y-\mu(\hat{\bm{\Theta}}_{k-1}))\right)^2 + \left(S_{\rho\lambda_{k-1}}((D \odot G_j)^\intercal(y-\mu(\hat{\bm{\Theta}}_{k-1})))\right)^2} \le (1-\rho)(2\lambda_k - \lambda_{k-1}),
\end{align*}
where $S_{\lambda}(\cdot)$ is the soft-tresholding function defined as
\begin{align*}
S_{\lambda}(a) = 
\begin{cases} 
a - \lambda & \text{if $ a > \lambda$} \\ 
0 & \text{if $|a| \le \lambda$} \\ 
a + \lambda & \text{if $a < -\lambda$}
\end{cases}.    
\end{align*}

\subsection{Prediction}\label{subsec:prediction}
Our proposed method to calculate prediction scores in individuals that were not used in training the models is presented in this section. In sparse regularized PQL estimation, we iteratively fit on a training set of size $n$ the working linear mixed model $$\tilde{\bm{Y}} = {\bm{X}}\hat{\bm{\Theta}} + \tilde{\bm{\bm{b}}} + \bm{\epsilon},$$ where $\hat{\bm{\Theta}} = \{ \hat{\Theta}_k \ne 0 | 1 \le k \le 2p + m + 1 \}$ is the set of non-null predictors, and $\bm{\epsilon}=g'(\bm{\mu})(\bm{y}-\bm{\mu}) \sim \mathcal{N}(0, \bm{W}^{-1})$, with $\bm{W} = \phi^{-1}\textrm{diag}\left\{ \frac{a_i}{\nu(\mu_i)[g'(\mu_i)^2]}\right\}$ the diagonal matrix containing weights for each observation. Let $\tilde{\bm{Y}}_s$ be the latent working vector in a testing set of $n_s$ individuals with predictor set ${\bm{X}}_s$. Similar to~\citep{Bhatnagar2020}, we assume that the marginal joint distribution of $\tilde{\bm{Y}}_s$ and $\tilde{\bm{Y}}$ is multivariate Normal :
\begin{align*}
\begin{bmatrix} \tilde{\bm{Y}}_s \\ \tilde{\bm{Y}}\end{bmatrix} \sim \mathcal{N}\left(\begin{bmatrix} \bm{X}_s\hat{\bm{\Theta}} \\ \bm{X}\hat{\bm{\Theta}}\end{bmatrix},\begin{bmatrix} \bm{\Sigma}_{11} & \bm{\Sigma}_{12} \\ \bm{\Sigma}_{21} & \bm{\Sigma}_{22} \end{bmatrix}\right),
\end{align*}
where $\bm{\Sigma}_{12}= \textrm{Cov}(\tilde{\bm{Y}}_s, \tilde{\bm{Y}}) = \hat\tau_g \bm{K}_{12} + \hat\tau_d \bm{K}^D_{12}$ is the sum of the $n_s \times n$ GSMs between the testing and training individuals, and $\bm{\Sigma}_{22}=\textrm{Var}(\tilde{\bm{Y}})=\bm{W}^{-1} + \hat\tau_g \bm{K}_{22} + \hat\tau_d \bm{K}^D_{22}$. It follows from standard normal theory that
\begin{align*}
\tilde{\bm{Y}}_s|\tilde{\bm{Y}}, \hat\phi,\hat{\bm\tau}, \hat{\bm{\Theta}}, \bm{X}, \bm{X}_s \sim \mathcal{N}\left(\bm{X}_s\hat{\bm{\Theta}} + \bm{\Sigma}_{12}\bm{\Sigma}_{22}^{-1}(\tilde{\bm{Y}}-\bm{X}\hat{\bm{\Theta}}), \bm{\Sigma}_{11}-\bm{\Sigma}_{12}\bm{\Sigma}_{22}^{-1}\bm{\Sigma}_{21}\right).
\end{align*}
The predictions are based on the conditional expectation $\E[\tilde{\bm{Y}}_s|\tilde{\bm{Y}}, \hat\phi,\hat{\bm\tau}, \hat{\bm{\Theta}}, \bm{X}, \bm{X}_s]$, that is
\begin{align*}
\hat{\bm{\mu}}_s &= g^{-1}\left(\E[\tilde{\bm{Y}}_s|\tilde{\bm{Y}}, \hat\phi,\hat{\bm\tau}, \hat{\bm{\Theta}}, \bm{X}, \bm{X}_s]\right) \nonumber \\ 
&= g^{-1}\left(\bm{X}_s\hat{\bm{\Theta}} + \bm{\Sigma}_{12}\bm{\Sigma}_{22}^{-1}(\tilde{\bm{Y}}-\bm{X}\hat{\bm{\Theta}})\right) \nonumber \\
&= g^{-1}\left(\bm{X}_s\hat{\bm{\Theta}} + \bm{\Sigma}_{12}\left(\bm{W}^{-1} + \bm{U\Lambda U}^\intercal\right)^{-1}(\tilde{\bm{Y}}-\bm{X}\hat{\bm{\Theta}})\right) ,
\end{align*}
where $g(\cdot)$ is the link function and ${\bm{U\Lambda U}}^\intercal$ is the spectral decomposition of the GSM for training subjects, with $\bm{U}$ the $n \times n$ matrix of eigenvectors. 

\section{Proximal Newton method}\label{appendix:2}
Defining the working vector $\tilde{\bm{Y}} = \bm{X\Theta}^{(t)} + \bm{U}\tilde{\bm{\delta}}^{(t+1)} + s_t\bm{\Delta}^{(t)}(\bm y - \bm\mu^{(t)})$ with suitable step size $s_t$, we can again rewrite the minimization problem as a WLS problem where
\begin{align}\label{eq:proxnewton_}
\bm{{\Theta}}^{(t+1)} &= \underset{\bm{{\Theta}}}{\textrm{argmin }} \left\{\frac{1}{2s_t} \left\|\bm{{\Theta}}- \left[\bm{X}^\intercal\bm{\Delta}^{-(t)}\bm{X}\right]^{-1}\bm{X}^\intercal\bm{\Delta}^{-(t)}\left(\bm{\tilde{Y}}^{(t)} - \bm{U}\bm{\tilde{\delta}}^{(t+1)}\right)\right\|_2^2 + g(\bm{\Theta})\right\} \nonumber \\
&= \underset{\bm{{\Theta}}}{\textrm{argmin }} \left\{\frac{1}{2s_t} \sum_{i=1}^n w_i\left(\tilde{Y}_i - \bm{X}_i\bm{\Theta} - \bm{U}_i\tilde{\bm{\delta}}^{(t+1)}\right)^2 + (1-\rho)\lambda\sum_j \|\beta_j, \gamma_j\|_2 + \rho\lambda\sum_j|\gamma_j|\right\},
\end{align}
with $w_i = \bm{\Delta}_{ii}^{-(t)}$. We use block coordinate descent and minimize \eqref{eq:proxnewton_} with respect to each component of $\bm{\Theta}=\left(\bm{\theta}^\intercal, \alpha^\intercal, \bm\beta^\intercal ,\bm{\gamma}^\intercal\right)^\intercal$. Suppose we have estimates $\tilde{\theta}_l$ for $l\ne j$, $\tilde{\bm{\beta}}$, $\tilde{\bm{\gamma}}$ and $\tilde{\bm{\delta}}$, it is straightforward to show that the updates for $\theta_j$ and $\alpha$ are given by
\begin{align*}
\tilde{\theta_j} &\leftarrow \frac{\sum_{i=1}^n w_i 
Z_{ij}\left(\tilde{Y}_i - \sum_{l\ne j}Z_{il}\tilde{\theta_l} - D_i \tilde{\alpha} - \bm{G}_i \tilde{\bm{\beta}} - (D_i \odot \bm{G}_i) \tilde{\bm{\gamma}}-\bm{U}_i\tilde{\bm{\delta}}\right)}{\sum_{i=1}^n w_i Z_{ij}^{2}}, \\
\tilde{\alpha} &\leftarrow \frac{\sum_{i=1}^n w_i
D_{i}\left(\tilde{Y}_i-\bm{Z}_i \tilde{\bm{\theta}} - \bm{G}_i \tilde{\bm{\beta}} - (D_i \odot \bm{G}_i) \tilde{\bm{\gamma}}-\bm{U}_i\tilde{\bm{\delta}}\right)}{\sum_{i=1}^n w_i D_{i}^{2}}.
\end{align*}
Denote the residual $r_{i;-j} = \tilde{Y}_i - \bm{Z}_{i}\tilde{\bm\theta} - D_i \tilde{\alpha} - \sum_{l\ne j}G_{il}\tilde{\beta_l} - \sum_{l\ne j}(D_i \odot G_{il}) \tilde{\gamma}_l-\bm{U}_i\tilde{\bm{\delta}}$. The subgradient equations for ${\beta}_j$ and ${\gamma}_j$ are equal to 
\begin{align*}
0 \in \begin{bmatrix}-\sum_{i=1}^n w_i 
G_{ij}\left(r_{i;-j}- {G}_{ij} \tilde{\beta_j} - (D_i \odot G_{ij}) \tilde{\gamma}_j\right) \\ -\sum_{i=1}^n w_i 
(D_i \odot G_{ij})\left(r_{i;-j}- {G}_{ij} \tilde{\beta_j} - (D_i \odot G_{ij}) \tilde{\gamma}_j\right) + \rho \lambda s_t \partial\|\tilde\gamma_j\|_1 \end{bmatrix} + (1-\rho)\lambda s_t \partial\|\tilde\beta_j, \tilde\gamma_j\|_2,  
\end{align*}
where we define the subgradients 
\begin{align*}
u \in \partial\|\tilde\gamma_j\|_1 = \begin{cases}
      [-1, 1] & \text{if $\tilde\gamma_j = 0$}\\
      \textrm{sign}(\tilde\gamma_j) & \text{if $\tilde\gamma_j \ne 0$}
    \end{cases};
\qquad \bm v \in \partial\|\tilde\beta_j, \tilde\gamma_j\|_2 = \begin{cases}
      \{\bm v|\ \|\bm v\|_2 \le 1\} & \text{if $\tilde\beta_j = \tilde\gamma_j = 0$}\\
      \frac{1}{\|\tilde\beta_j, \tilde\gamma_j\|_2}\begin{bmatrix}\tilde\beta_j \\ \tilde\gamma_j\end{bmatrix} & \text{otherwise}
    \end{cases}.
\end{align*}
(1) The case $\tilde\beta_j = \tilde\gamma_j = 0$ implies 
\begin{align*}
\begin{bmatrix} \sum_{i=1}^n w_i 
G_{ij}r_{i;-j}\\ \sum_{i=1}^n w_i 
(D_i\odot G_{ij})r_{i;-j}- \rho \lambda s_tu \end{bmatrix} = (1-\rho)\lambda s_t \bm v.
\end{align*}
Since $\|\bm v\|_2 \le 1$, equality of the constraint holds as long as
\begin{align*}
\left(\sum_{i=1}^n w_i 
G_{ij}r_{i;-j}\right)^2 + \left(\sum_{i=1}^n w_i 
(D_i\odot G_{ij})r_{i;-j}- \rho\lambda s_t u\right)^2 \le ((1-\rho)\lambda s_t)^2.
\end{align*}
Since $u\in[-1, 1]$, a necessary and sufficient condition for $\tilde\beta_j = \tilde\gamma_j = 0$ being a solution is
\begin{align}\label{ineq:2}
\left(\sum_{i=1}^n w_i 
G_{ij}r_{i;-j}\right)^2 + \left(S_{\rho\lambda s_t} \left(\sum_{i=1}^n w_i 
(D_i\odot G_{ij})r_{i;-j}\right)\right)^2 \le ((1-\rho)\lambda s_t)^2,
\end{align}
where $S_{\lambda}(\cdot)$ is the soft-tresholding function defined as
\begin{align*}
S_{\lambda}(a) = 
\begin{cases} 
a - \lambda & \text{if $ a > \lambda$} \\ 
0 & \text{if $|a| \le \lambda$} \\ 
a + \lambda & \text{if $a < -\lambda$}
\end{cases}.    
\end{align*}
(2) The case $(\tilde\beta_j, \tilde\gamma_j)^\intercal \ne \bm 0$ implies
\begin{align}\label{eq:sub3}
\begin{bmatrix} \sum_{i=1}^n w_i 
G_{ij}(r_{i;-j} - D_iG_{ij}\tilde{\gamma}_j)\\ \sum_{i=1}^n w_i 
(D_i\odot G_{ij})(r_{i;-j} - G_{ij}\tilde{\beta}_j) - \rho\lambda s_tu \end{bmatrix} = \left(\begin{bmatrix}\sum_{i=1}^n w_i 
G_{ij}^2 & 0 \\ 0 & \sum_{i=1}^n w_i 
(D_i\odot G_{ij})^2\end{bmatrix} + \frac{(1-\rho)\lambda s_t}{\sqrt{\tilde\beta_j^2 + \tilde\gamma_j^2}}\bm{I}_2 \right) \begin{bmatrix} \tilde\beta_j \\ \tilde\gamma_j \end{bmatrix}.
\end{align}
We have that $\tilde\gamma_j = 0$ if $|\sum_{i=1}^n w_i 
(D_i\odot G_{ij})(r_{i;-j}-G_{ij}\tilde{\beta}_j)| \le \rho\lambda s_t$ since $u \in [-1, 1]$. 
This implies that
\begin{align*}
 \sum_{i=1}^n w_i G_{ij}r_{i;-j}= \left(\sum_{i=1}^n w_i  G_{ij}^2 + (1-\rho)\frac{\lambda s_t}{|\tilde\beta_j|}\right)\tilde\beta_j,
\end{align*}
with the solution being equal to
\begin{align*}
    \tilde\beta_j = \frac{S_{(1-\rho)\lambda s_t}(\sum_{i=1}^n w_i G_{ij}r_{i;-j})}{\sum_{i=1}^n w_i  G_{ij}^2}.
\end{align*}
There is no closed-form solution for \eqref{eq:sub3} if both $\tilde\gamma_j$ and $\tilde\beta_j$ are non-null. In this case, we can replace \eqref{eq:proxnewton_} by a surrogate objective function using a majorization-minorization algorithm~\citep{Wu2008}. From the concavity of the $\ell_2$ norm $\|\beta_j, \gamma_j\|_2 = \sqrt{\beta_j^2+\gamma_j^2}$, we have the following inequality
\begin{align*}
\|\beta_j, \gamma_j\|_2 \le \|\beta_j^{(t)}, \gamma_j^{(t)}\|_2 + \frac{1}{2\|\beta_j^{(t)}, \gamma_j^{(t)}\|_2}(\|\beta_j, \gamma_j\|_2^2 - \|\beta_j^{(t)}, \gamma_j^{(t)}\|_2^2),
\end{align*}
from where we derive the majorization-minimization iterative step
\begin{align*}
\bm{{\Theta}}^{(t+1)} &= \underset{\bm{{\Theta}}}{\textrm{argmin }} \left\{\frac{1}{2s_t} \sum_{i=1}^n w_i\left(\tilde{Y}_i^{(t)} - \bm{X}_i\bm{\Theta} - \bm{U}_i\bm{\delta}^{(t+1)}\right)^2 + (1-\rho)\lambda \sum_j\frac{\|\beta_j, \gamma_j\|_2^2}{2\|\beta_j^{(t)}, \gamma_j^{(t)}\|_2} + \rho\lambda \sum_j |\gamma_j|\right\}.
\end{align*}
Using cyclic coordinate descent, the updates for $\beta_j$ and $\gamma_j$ are given by
\begin{align*}
\tilde{\beta_j} &\leftarrow \frac{\sum_{i=1}^n w_i 
G_{ij}\left(r_{i;-j}-D_iG_{ij}\tilde{\gamma}_j\right)}{\sum_{i=1}^n w_i G_{ij}^{2} + (1-\rho)\lambda \tilde{s_t}}, \\
\tilde{\gamma} &\leftarrow \frac{S_{\rho\lambda {s}_t}\left(\sum_{i=1}^n w_i 
D_iG_{ij}(r_{i;-j}-G_{ij}\tilde{\beta}_j)\right)}{\sum_{i=1}^n w_i (D_iG_{ij})^{2} + (1-\rho)\lambda \tilde{s_t}},
\end{align*}
where we defined $\tilde{s}_t=s_t/\|\beta_j^{(t)}, \gamma_j^{(t)}\|_2$. Algorithm 1 below summarizes our block coordinate descent (BCD) procedure to obtain regularized estimates for the fixed effects vector $\bm{\Theta}=\left(\bm{\theta}^\intercal, \alpha^\intercal, \bm\beta^\intercal ,\bm{\gamma}^\intercal\right)^\intercal$.

\scriptsize
\begin{algorithm}[H]
\DontPrintSemicolon
  
\KwInput{$y, \bm{X}=[\bm{Z}\ \bm{D}\ \bm{G}\ (\bm{D} \odot \bm{G})]$}
  \KwOutput{$\hat{\bm{\theta}}, \hat{\alpha}, \hat{\bm{\beta}}, \hat{\bm{\gamma}}$}
    
  Estimate $\tau_g, \tau_d$ and $\phi$ under the null model (i.e. $\bm{\beta}=\bm{\gamma}=\bm{0}$) using the AI-REML algorithm;
  
  Given $\hat\tau_g, \hat\tau_d$ and  $\hat\phi$, perform spectral decomposition of the random effects covariance matrix $ \hat\tau_g \bm{K} + \hat\tau_d \bm{K}^D = \bm{U}\bm{\Lambda}\bm{U}^\intercal$;
  
   Initialize $\bm{\Theta}^{(0)}=({\bm{\theta}^{(0)}}^\intercal, {\alpha^{(0)}}^\intercal, {\bm{\beta}^{(0)}}^\intercal, {\bm{\gamma}^{(0)}}^\intercal)^\intercal \textrm{ and } \tilde{\bm{\delta}}^{(0)}$ ;
  
  \For{$\lambda=\lambda_1,\lambda_2,...$}{
  \For{t=0,1,... until convergence}{
    
    Select a suitable step size $s_t$;
    
    Update $\bm{\mu}^{(t)} \leftarrow g^{-1}(\bm{X}\bm{\Theta}^{(t)} + \bm{U}\tilde{\bm{\delta}}^{(t)})$, $\bm{\Delta}^{(t)} \leftarrow \textrm{diag}(g'(\bm{\mu}^{(t)}))$ and $w_i \leftarrow \bm{\Delta}_{ii}^{-(t)}$ for $i=1,...,n$;
    
    Update $\tilde{\bm{Y}} \leftarrow \bm{X}\bm{\Theta}^{(t)} + \bm{U}\tilde{\bm{\delta}}^{(t)} + s_t\bm{\Delta}^{(t)}(\bm y - \bm\mu^{(t)})$;
    
    \underline{/*\textbf{ Inner loop to estimate $\tilde{\bm{\delta}}$}}
        	
	\For{j=1,...,n until convergence}     
        { 
            \begin{align*}
            \tilde{\delta_j}^{(t+1)} \leftarrow \frac{\sum_{i=1}^n w_iU_{ij}\left(\tilde{Y}_i- \bm{X}_i\bm{\Theta}^{(t)} - \sum_{l\ne j}U_{il}\tilde{\delta_l}\right)}{\sum_{i=1}^n w_i U_{ij}^{2} + \hat{\phi}\Lambda_j^{-1}};  
            \end{align*}
        }
                
     
     Update $\bm{\mu}^{(t)} \leftarrow g^{-1}(\bm{X}\bm{\Theta}^{(t)} + \bm{U}\tilde{\bm{\delta}}^{(t+1)})$;
     
     Update $\tilde{\bm{Y}} \leftarrow \bm{X}\bm{\Theta}^{(t)} + \bm{U}\tilde{\bm{\delta}}^{(t+1)} + s_t\bm{\Delta}^{(t)}(\bm y - \bm\mu^{(t)})$;
     
     \underline{/*\textbf{ Inner loop to estimate $\bm{\Theta}^{(t+1)}$}}
  
	\For{k=1,...,m until convergence}     
        { 
            \begin{align*}
            \tilde{\theta_k} &\leftarrow \frac{\sum_{i=1}^n w_i 
            Z_{ij}\left(\tilde{Y}_i - \sum_{l\ne k}Z_{il}\tilde{\theta_l} - D_i \tilde{\alpha} - \bm{G}_i \tilde{\bm{\beta}} - (D_i \odot \bm{G}_i) \tilde{\bm{\gamma}}-\bm{U}_i\tilde{\bm{\delta}}\right)}{\sum_{i=1}^n w_i Z_{ik}^{2}}, \\
            \tilde{\alpha} &\leftarrow \frac{\sum_{i=1}^n w_i
            D_{i}\left(\tilde{Y}_i-\bm{Z}_i \tilde{\bm{\theta}} - \bm{G}_i \tilde{\bm{\beta}} - (D_i \odot \bm{G}_i) \tilde{\bm{\gamma}}-\bm{U}_i\tilde{\bm{\delta}}\right)}{\sum_{i=1}^n w_i D_{i}^{2}};
            \end{align*}
         }  
         \For{j=1,...,p until convergence}     
        {
            Compute $r_{i;-j} = \tilde{Y}_i - \bm{Z}_{i}\tilde{\bm\theta} - D_i \tilde{\alpha} - \sum_{l\ne j}G_{il}\tilde{\beta_l} - \sum_{l\ne j}(D_i \odot G_{il}) \tilde{\gamma}_l-\bm{U}_i\tilde{\bm{\delta}}$;

            If $|\sum_{i=1}^n w_i 
            (D_i\odot G_{ij})(r_{i;-j}-G_{ij}\tilde{\beta}_j)| \le \lambda s_t$ then set $$\tilde\gamma_j\leftarrow 0 \text{ and }\tilde\beta_j \leftarrow \frac{S_{\lambda s_t}(\sum_{i=1}^n w_i G_{ij}r_{i;-j})}{\sum_{i=1}^n w_i  G_{ij}^2};$$

            Else then set
            \begin{align*}
            \tilde{\beta_j} &\leftarrow \frac{\sum_{i=1}^n w_i 
            G_{ij}\left(r_{i;-j}-D_iG_{ij}\tilde{\gamma}_j\right)}{\sum_{i=1}^n w_i G_{ij}^{2} + \lambda \tilde{s_t}}, \\
            \tilde{\gamma} &\leftarrow \frac{S_{\lambda \tilde{s}_t}\left(\sum_{i=1}^n w_i 
            D_iG_{ij}(r_{i;-j}-G_{ij}\tilde{\beta}_j)\right)}{\sum_{i=1}^n w_i (D_iG_{ij})^{2} + \lambda \tilde{s_t}},
            \end{align*}
            where $\tilde{s}_t=s_t/\|\beta_j^{(t)}, \gamma_j^{(t)}\|_2$.

        }
    }
 }
\caption{BCD algorithm to minimize the PQL loss function of the GEI model \eqref{eq:objfunc2} with mixed lasso and group lasso penalties for GLMMs.}
\end{algorithm}

\end{appendix}
\end{document}